\documentclass[useAMS,usenatbib]{mn2e}
\usepackage{amssymb}
\usepackage{amsmath}
\usepackage{graphicx}
\usepackage{lscape}
\usepackage{color}
\usepackage{subfigure}

\def\h2{\rm{H_2}}
\def\fh2{f_{\rm{H_2}}}

\def\Sh2{\Sigma_{\h2}}
\def\sgas{\Sigma_{\rm{gas}}}

\def\ms{M_{\odot}}
\def\Hs{M_{\h2}/M_*}
\def\HIs{M_{\rm{HI}}/M_*}

\title[Radial Gradients in SAMs of Disk Galaxy Formation]
{Star Formation and Metallicity Gradients in Semi-analytic Models of Disk Galaxy Formation}

\author[Fu et al.]{Jian Fu$^{1,2}$ \thanks{E-mail: fujian@mpa-garching.mpg.de}, Guinevere Kauffmann$^{1}$, Mei-ling Huang$^1$, Robert M. Yates$^1$,
\newauthor Sean Moran$^3$, Timothy M. Heckman$^4$, Romeel Dav\'e$^5$, Qi Guo$^{6,7}$,
\newauthor Bruno M. B. Henriques$^1$ \\
$^1$Max-Planck-Institut f\"ur Astrophysik, D-85741 Garching, Germany  \\
$^2$Key Laboratory for Research in Galaxies and Cosmology, Shanghai Astronomical Observatory, CAS,\\ 80 Nandan Rd., Shanghai, 200030, China\\
$^3$Harvard-Smithsonian Center for Astrophysics, 60 Garden Street, Cambridge, MA 02138, USA \\
$^4$Department of Physics and Astronomy, The Johns Hopkins University, MD21218, Baltimore, USA\\
$^5$Astronomy Department, University of Arizona, AZ85721, Tucson, USA\\
$^6$Institute for Computational Cosmology, Department of Physics, University of Durham, South Road, Durham, DH1 3LE, UK\\
$^7$National Astronomical Observatories, CAS, Beijing 100012, China
}

\begin{document}

\maketitle

\begin{abstract}

We have updated our radially-resolved semi-analytic models of galaxy formation, which track both the atomic and molecular gas phases of the interstellar medium. The models are adapted from those of Guo et al. (2011) using similar methodology as in Fu et al. (2010) and are run on halo merger trees from the Millennium and Millennium II simulations with the following main changes: (1) We adopt a simple star formation law $\Sigma_{\rm SFR} \propto \Sh2$. (2) We inject the heavy elements produced by supernovae directly into the halo hot gas, instead of first mixing them with the cold gas in the disk. (3) We include radial gas inflows in disks using a model of the form $v_{\rm inflow} = \alpha r$. The models are used to study the radial profiles of star formation rate and gas-phase metallicity in present-day galaxies. The surface density profiles of molecular gas in $L_*$ galaxies place strong constraints on inflow velocities, favouring models where $v_{\rm inflow}\sim$ 7 km/s at a galactocentric radius of 10 kpc. Radial gas inflow has little influence on gas-phase and stellar metallicity
gradients, which are affected much more strongly by the fraction of metals that are directly injected into the halo gas, rather than mixed with the cold gas. Metals ejected out of the galaxy in early epochs result in late infall of pre-enriched gas and {\em flatter} present-day gas-phase metallicity gradients. A prescription in which $80\%$ of the metals are injected into the halo gas results in good fits to the flat observed metallicity gradients in galaxies with stellar masses greater than $10^{10} M_{\odot}$, as well as the relations between gas-phase metallicity and specific star formation rate in the outer parts of galactic disks. We examine the correlation between gas-phase metallicity gradient and global galaxy properties, finding that it is most strongly correlated with the bulge-to-total ($B/T$) ratio of the galaxy. This is because gas is consumed when the bulge forms during galaxy mergers, and the gas-phase metallicity gradient is then set by
newly-accreted gas.

\end{abstract}


\section{Introduction}

In recent years, observations of radially resolved profiles of atomic gas, molecular gas and star formation in representative samples nearby galaxies (e.g THINGS for HI profiles, Walter et al. 2008; HERACLES for $\h2$ profiles, Leroy et al. 2009) have motivated galaxy formation theorists to include the physics of the atomic-to-molecular gas transition in their models. Some of these models make predictions for the global atomic and molecular gas content of galaxies (e.g. Obreschkow et al. 2009; Lagos et al. 2011), while others attempt to model the radial structure of the gas in more detail (e.g. Robertson \& Kravtsov 2008; Fu et al. 2010; Power, Baugh \& Lacey 2010; Feldmann, Hernandez \& Gnedin 2012). Radial abundance gradients also place important constraints on disk formation models, particularly on how supernova feedback (SN feedback) processes eject metals into the surrounding gas and how these metals are mixed throughout the disk as the galaxy evolves. A successful model of disk galaxy formation should be able to reproduce the metallicity gradients in galaxies along with radial density profiles of old stars, young stars and gas.

Observational data on metal abundance gradients in galaxies remains rather sparse. Many studies focus only on one galaxy or on a handful of galaxies (e.g Rudolph et al. 2006 for the Milky Way; Magrini et al. 2007 for M33; Bresolin et al. 2009 for M83). For many years, the Zaritsky, Kennicutt \& Huchra (1994) study of gas-phase metallicity gradients derived for 159 HII regions in 14 spiral galaxies in combination with published data for another 25 galaxies, has remained the standard reference in the field. More recently, Moustakas et al. (2010) published metallicity gradients for 65 galaxies from the SINGS survey and Moran et al. (2012; hereafter Moran12) analyzed gradients for 174 galaxies with atomic and molecular gas mass measurements from the GASS and COLD GASS surveys (Catinella et al. 2010; Saintonge et al. 2011a). These new observations make it possible to carry out a statistical comparison with model predictions for the first time.

There is a long history of models of the radial structure and properties of disk galaxies in the literature, beginning with Tinsley \& Larson (1978) who implemented a model for chemical evolution into the dynamical collapse calculations for gas clouds with rotation and axial symmetry introduced by Larson (1976). Similar modelling efforts were later undertaken by many others (e.g. Matteucci \& Francois 1989; Kauffmann 1996; Chiappini, Matteucci \& Gratton 1997; Dalcanton, Spergel \& Summers 1997; Avila-Reese, Firmani, \& Hern{\'a}ndez 1998; van den Bosch 1998; Prantzos 1999; Dutton et al. 2007; Fu et al. 2009; Yin et al. 2009; Cook et al. 2010; Fu et al. 2010). The models can be arranged in increasing order of complexity, from those that use simple parameterized formulae to describe the infall of gas onto the disk, to those that are embedded within high resolution N-body simulations of structure formation in a $\Lambda$CDM cosmology and where gas infall rates are determined by explicitly following the growth of each dark matter halo in the simulation and then calculating the rate at which gas will cool as a function of time.

In order to model the radial distributions of stars and gas in the disk, processes such as star formation, SN feedback and chemical enrichment need to be treated in a radially resolved fashion. In previous work (Fu et al. 2010, hereafter Fu10), we developed models in which each galaxy disk was divided into a series of concentric rings. We adopted two different prescriptions to partition the cold gas into atomic and molecular components: one is adapted from the models of Krumholz, McKee \& Tumlinson (2009), in which the $\h2$ fraction is parameterized as a function of local cold gas surface density and gas-phase metallicity. The other prescription is empirically-based and proposes that the $\h2$ fraction is related to the interstellar pressure (Elmegreen 1989 \& 1993; Blitz \& Rosolowsky 2004 \& 2006). The Fu10 semi-analytic model is based on the version of the L-Galaxies code described in De Lucia \& Blaizot (2007) (hereafter DLB07) implemented on the halo merger trees of the Millennium Simulation outputs (Springel et al. 2005). As discussed in the paper, the model reproduces the radial profiles of stellar mass, HI, $\h2$, and star formation in $L_*$ galaxies. However, the model produces a stellar mass function that is too steep at the faint end and the gas fractions of low mass galaxies do not agree with observations, which is a problem inherit from DLB07 model.

The analysis presented in this paper is based on the new version of the L-Galaxies code described in Guo et al. (2011) (hereafter Guo11), which runs on the halo merger trees of both the Millennium Simulation (Springel et al. 2005) and on the 125 times higher resolution Millennium II Simulation (Boylan-Kolchin et al. 2009). This allows us to study the formation and evolution of galaxies with masses ranging from those of dwarfs to the most massive cD galaxies. Based on the new models, the main purpose of this paper is to study the radial profiles of the gas-phase metallicity and star formation rate surface density of disk galaxies at $z=0$ and compare them with the recent observational results, especially those from Moran12.

This paper is organized as follows. In Section 2, we briefly describe the N-body simulation and the L-Galaxies semi-analytic models, and then we will give the changes to the semi-analytic model with respect to Fu10 and Guo11. We discuss how we normalize the free parameters in our models using gas profile data from Leroy et al. (2008) for a small sample of nearby disks, along with the stellar and gas mass functions of nearby galaxies at $z=0$. In Section 3, we present stellar, HI and $\h2$ mass functions at $z=0$ and compare them to observations. In Section 4, we analyze how the star formation rate surface density profiles of galaxies depend on stellar mass. In Section 5, we study how the radial profiles of gas-phase metallicity depend on stellar mass, bulge-to-total ratio, stellar surface density and specific star formation rate. We also look at the relations between metallicity and star formation in the inner and outer disks of galaxies. The model results presented in Sections 4 and 5 are compared to observational results from Moran12. Finally, in Section 6, we summarize our results and discuss avenues for future work.

\section{The N-body simulation and semi-analytic models}

In this section, we will briefly introduce the Millennium and Millennium II simulations, the physical processes in L-Galaxies semi-analytic galaxy formation models, the methods of treating the radial profiles and the conversion of atomic to molecular gas in the semi-analytic model framework.

\subsection{The N-body simulations} \label{chap:ms}

The L-Galaxies semi-analytic models of galaxy formation are run on two very large N-Body simulations: the Millennium Simulation (hereafter MS, Springel et al. 2005) and Millennium-II Simulation (hereafter MS-II, Boylan-Kolchin et al. 2009). Both simulations adopt a $\rm{\Lambda}$CDM cosmogony with parameters $\Omega_\Lambda=0.75, ~\Omega_m=0.25,~\Omega_{\rm{baryon}}=0.045,~\sigma_8=0.9$ and $h=0.73$, and they track $2160^3\approx10^{10}$ particles from $z=127$ to $z=0$. The main difference between the two simulations is the resolution. The particle mass in MS is $8.6\times10^{8}\ms~h^{-1}$ and the periodic box is 500 Mpc~$h^{-1}$ on a side; the particle mass in MS-II is $6.8\times10^{6}\ms~h^{-1}$ and the size of the periodic box is 100 Mpc~$h^{-1}$, which means the resolution of MS-II is 125 times higher than MS. Since the smallest halo traced in both simulations has 20 particles, MS can be used to study the formation and evolution for Milky Way-sized or larger galaxies, while MS-II is more appropriate to study dwarf galaxies. For our models of the cold gas components in galaxies, the MS is suitable to study atomic and molecular gas for disk galaxies with stellar mass $M_*\gtrsim10^{10}\ms$ and MS-II is useful to study the gas components in present-day dwarf galaxies and in the high-redshift progenitors of present-day $L_*$ galaxies. Another difference between the two simulations is the number of output snapshots: the last 60 snapshots for MS and MS-II are identical, but MS-II has four more snapshots at very high redshift to increase the time resolution in early-forming first structures.

\subsection{The galaxy formation models} \label{chap:model}

The L-Galaxies is the semi-analytic model of galaxy formation developed by Munich group (Kauffmann et al. 1999; Springel et al. 2001; Croton et al. 2006; DLB07; Guo11; Guo et al. 2013). The model tracks various physical processes on the halo merger trees of MS and MS-II: re-ionization, gas infall and cooling, star formation and metal production, SN feedback, ram-pressure stripping of gas in satellite galaxies, tidal disruption of satellites, galaxy mergers, bulge formation, black hole growth and AGN feedback.

\begin{figure}
\centering
 \includegraphics[angle=0,scale=0.55]{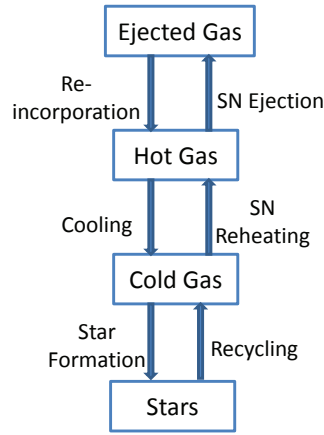}
 \caption
 {Illustration of how mass is exchanged between different phases in model galaxies in response to the physical processes treated by the model.
 }\label{fig:processes}
\end{figure}

The L-Galaxies model tracks four phases: stars, interstellar cold gas, hot gaseous haloes (hot gas) and ejecta reservoirs (ejected gas). In Fig. \ref{fig:processes}, we illustrate how mass is exchanged between different phases when different physical processes operate. The exchange of metals follows the exchange of the gas. All metals are fully mixed with the gas in one timestep. In Appendix \ref{chap:appendix}, we briefly summarize the physics and the equations of the processes in Fig. \ref{fig:processes}, and the descriptions of bulge formation and mergers will be described in Sec. \ref{chap:bulgeprofiles}. The reader is also referred to Section 3 of Guo11 for a more detailed discussion of how the physical processes in the model are calculated.

The cosmological parameters for both MS and MS-II and the semi-analytic models in Guo11 and previous papers are from WMAP1 (Wilkinson Microwave Anisotropy Probe first-year results, Spergel et al. 2003). To update the cosmology, Guo et al. (2013, hereafter Guo13) adopt the technique described in Angulo \& White (2010) to rescale the growth of structure to be appropriate for WMAP7 parameters ($\Omega_\Lambda=0.728, ~\Omega_m=0.272,~\Omega_{\rm{baryon}}=0.0454, ~\sigma_8=0.807$ and $h=0.704$; Komatsu et al. 2011). In this paper, we make use of these rescaled models, but we note that differences for the quantities explored in this paper are negligible compared to systematic uncertainties in our treatment of the various gas-physical processes.

The main changes made in this paper with respect to Guo11 are the following: (1) We include the methodology to track gas and stellar components in radial rings described in Fu10. (2) We include the prescriptions for the transition between atomic gas and molecular gas described in Fu10. (3) We abandon the two-regime star formation model in Fu10 and adopt a simpler prescription where the star formation surface density is always proportional to the molecular gas surface density (e.g. Leroy et al. 2008; Bigiel et al. 2008; Schruba et al. 2011). (4) We allow a fraction of heavy elements to be mixed directly with the hot gas in the halo. We describe these modifications in more detail below.

\subsection{Treatment of the radial distribution of gas and star in disks}

In the ``standard'' L-Galaxies recipes (DLB07 \& Guo11), each galaxy disk is treated as a single zone. However, these single zone recipes are not suitable to study the transition of atomic gas to molecular gas, since molecular clouds form where the {\em local} density is high enough for molecules to be shielded from the surrounding ionizing radiation. Single zone models are also unable to predict metallicity gradients. The Fu10 models are based on the DLB07 models, except that we divide each model galaxy disk into a set of radial concentric rings to study the radial distribution of star and gas components on galaxy disk. We trace the physical processes related to the galaxy disks in each ring in each timestep. Thus, our model can predict the surface density profiles of stars, gas and star formation rate, and also the radial profiles of the stellar and gas-phase metallicity.

In Fu10, we divide each galaxy disk into 30 rings, and the radius of the rings is given by a geometric series. In this paper, we divide each disk into 12 rings instead of 30 rings. This is done to decrease the computational memory requirements based on MS-II halo merger trees. The ring radii are given by the geometric series
\begin{equation}\label{eq:ri}
{r_i} = 0.44 \times {1.5^i}~[h^{-1} \rm{kpc}]~(i=1,2...12)
\end{equation}
The radius for the innermost ring in Eq. (\ref{eq:ri}) is about 0.9 kpc, and the radius for the outermost ring is about 80 kpc. As we mentioned in Fu10, the radial profiles are insensitive to the precise scheme, if the adopted number of rings is sufficiently large.
This sub-division of the disk in Eq. (\ref{eq:ri}) is sufficient to study both small and large galaxies.

\subsubsection{The radial distribution of gas cooling onto the galaxy disks and inside-out disk growth}

To model the radial distribution of stars and gas in the disk of the galaxy, we adopt the same prescription described in Fu10, in which the surface density profile of newly infalling cold gas at each timestep has exponential form with a uniform metallicity (see Fig. \ref{fig:zhot} and the discussion in last section on the pre-enriched gas accretion). The scale length of the exponential infalling profile is given by $r_{\rm{infall}}=r_{\rm{d}}=\left(\lambda /\sqrt 2\right)r_{\rm{vir}}$, and this equation assumes that angular momentum is conserved during gas cooling and infall (Mo, Mao \& White 1998). In our simple scheme, the profile of the newly infalling gas is directly superposed onto the pre-existing gas profile from the previous timestep.

Since the disk and halo size is smaller and compact at high redshift, the scale length of the infalling gas is smaller at higher redshift. The size of the galaxy disk grows with time (see Figure 1 in Fu10). This is the so-called ``inside-out'' disk growth paradigm, that has been incorporated in many disk formation models (e.g Kauffmann et al. 1996; Dalcanton, Spergel, \& Summers 1997; Avila-Reese, Firmani, \& Hern{\'a}ndez 1998; Dutton 2009; Fu et al. 2009; Pilkington et al. 2012).

\subsubsection{The radial distribution of bulge formation and galaxy mergers} \label{chap:bulgeprofiles}

The DLB07 model does not include a prescription for bulge sizes and the Fu10 model did not attempt to calculate the radial distribution of bulge stars. The Guo11 models include a method for calculating the sizes of bulges produced during galaxy mergers and by disc instabilities.

The size of the bulge after a merger is calculated using the following equation
\begin{equation}\label{eq:starburstbulge}
C\frac{m_{\rm new}^2}{r_{\rm{new}}} = C\frac{m_1^2}{r_1} + C\frac{m_2^2}{r_2} + \alpha\frac{{{m_1}{m_2}}}{{{r_1} + {r_2}}}
\end{equation}
where $C$ parametrizes the binding energy of the galaxy and $\alpha$ parametrizes the effective interaction energy deposited in the stellar components. In Guo11, $C=0.5$ and $\alpha=0.5$ are adopted for galaxy mergers. $m_{\rm new}$ and $r_{\rm new}$ in Eq. (\ref{eq:starburstbulge}) are the mass and half mass radius of the newly formed bulge. For major mergers, in which the ratio of the total (stellar+gas) mass of the satellite to the central galaxy is larger than 0.3, both disks are destroyed. The stars in both progenitor galaxies and the stars formed in the merger-induced starburst become bulge stars. In this case, $m_1$ is sum of the stellar mass of the two progenitor galaxies and $m_2$ is the sum of the stellar mass converted from cold gas in the major merger starburst (see later); $r_1$ and $r_2$ are the corresponding half-mass radii.

For a minor merger ($m_{\rm sat}/m_{\rm cen}<0.3$), the stellar component of the smaller satellite galaxy is added to the bulge of the central galaxy, while the cold gas component of the satellite is added to the disk of the central galaxy. $m_1$ in Eq. (\ref{eq:starburstbulge}) is the bulge mass of the central galaxy before the merger and $m_2$ is the stellar mass (bulge star+disk star) of the satellite galaxy; $r_1$ and $r_2$ are the corresponding half mass radii.

The main new prescription implemented in this paper concerns the radial profile of gas added to the central galaxy from the satellite before the starburst occurs. The Fu10 code directly superposes the gas radial profiles of two gas disks together and then processes the starburst. In this paper, we treat the accreted gas from the satellite in the same way as gas cooling from the halo; the scale length of the accreted gas is determined from the spin parameter and virial radius of the halo of the central galaxy using the equation $r_{\rm infall}=(\lambda/\sqrt{2})r_{\rm vir}$.

The DLB07 and Guo11 codes use the ``collisional starburst model'' introduced by Somerville, Primack \& Faber (2001). In both major and minor mergers, the mass of stars formed in the starburst is $m_*=e_{\rm burst}m_{\rm gas}$, with $e_{\rm burst}$ given by the equation
\begin{equation}\label{eq:estarburst}
e_{\rm burst}=\beta _{\rm{burst}}\left(m_{{\rm{sat}}}/m_{\rm{cen}}\right)^{\alpha_{\rm{burst}}}
\end{equation}
with $\alpha_{\rm{burst}}=0.7$ and $\beta _{\rm{burst}}=0.56$.

In our model, we simply calculate the surface mass density of stars formed in the burst in ring number $i$ as
\begin{equation}\label{eq:starburst}
\Sigma_{*, i}=e_{\rm burst}\Sigma_{{\rm gas}{,i}}
\end{equation}
using Eq. (\ref{eq:estarburst}) to define $e_{\rm burst}$. For starbursts occurring during major mergers, the stars formed in all
the rings become bulge stars.

This is admittedly not a very realistic description of what happens in an actual merger-induced starburst, but since we mainly study quiescent disk galaxies in this paper, this will not concern us. The other update we make with respect to the Guo11 models during mergers is to calculate the half mass radii directly from the stellar and gas profiles, rather than to assume an exponential form.

Bulges also form through secular evolution. When $v_{\rm max}<\sqrt{G m_{*,\rm disk}/r_{*,\rm disk}}$, the disk is considered unstable. A mass $\delta m_*$ from the inner part of the disk is transferred to the bulge
\begin{equation}\label{eq:deltam}
\delta m_*=m_{*,\rm_{disk}}-\frac{r_{*,\rm disk}v_{\rm max}^2}G
\end{equation}
where $m_{*,\rm disk}$ and $r_{*,\rm disk}$ are the mass and scale length of the stellar disk. If this material forms a new bulge, it is assumed to have a half-mass radius $\delta r$ equal to the outer radius of the region. If there is already a bulge in the galaxy, then the resulting bulge is calculated using
\begin{equation}\label{eq:instablebulge}
0.5\frac{m_{\rm new}^2}{r_{\rm{new}}} =0.5 \frac{m_1^2}{r_1} +0.5 \frac{m_2^2}{r_2} + 2.0\frac{{{m_1}{m_2}}}{{{r_1} + {r_2}}}
\end{equation}
where $m_1$ and $r_1$ are the mass and half mass radius of the existing bulge; $m_2=\delta m_*$ and $r_2=\delta r$ are the mass and radius of the transferred region. Once again we have updated this prescription by calculating the radius $\delta r$ directly from the stellar radial distribution of the disk rather than by assuming an exponential stellar disk profile as in Guo11.

As in Guo11, we assume a Jaffe profile (Jaffe 1983) for the bulge of the form
\begin{equation}\label{eq:jaffe}
\rho_{\rm bulge} \left( r \right) = \frac{{{m_{{\rm{bulge}}}}}}{{4\pi r_{\rm{b}}^3}}{\left( {\frac{r}{{{r_{\rm b}}}}} \right)^{ - 2}}{\left( {1 + \frac{r}{{{r_{\rm{b}}}}}} \right)^{-2}}
\end{equation}
in which $m_{\rm bulge}$ is the bulge mass and $r_{\rm b}$ is the half mass radius of the galaxy bulge. The stellar mass in a given radial ring is the combined mass of disk and bulge stars in that ring:
\begin{multline}\label{eq:stellarmassr}
 m_*=m_{*, \rm disk}+4\pi \int_{r_{{\rm{in}}}}^{{r_{{\rm{out}}}}}{{\rho_{\rm{bulge}}}\left(r\right)}{r^2}dr \\
 =m_{*, \rm disk}+m_{\rm bulge}\left[\left(1+\frac{r_{\rm in}}{r_{\rm b}}\right)^{-1}-\left(1+\frac{r_{\rm out}}{r_{\rm b}}\right)^{-1}\right]
\end{multline}
where $m_{\rm *,disk}$ is the stellar mass of the disk component in the ring, and $r_{\rm in}$, $r_{\rm out}$ are the inner and outer radii of the radial ring. We note that the stellar surface density profiles presented in the next sections include both bulge and disk components.

\subsubsection{Radial gas inflows in disks} \label{chap:gasflow}

In Fu10, we showed that a star formation law of form $\Sigma_{\rm SFR}\propto \Sh2$ was problematic in that it led to cold gas surface density profiles that were too shallow to match observations (see Sec. 3.4 \& Fig. 3 in Fu10). In that paper, we introduced the following two tweaks to solve the problem: (i) we adopted a star formation law in regions of the disk with $f_{\h2}<0.5$ of the form $\Sigma _{{\rm{SFR}}}\propto\Sigma_{{\rm{gas}}}^2$. (ii) we assumed that the SN reheating efficiency was inversely proportional to gas surface density ($\Delta m_{\rm{reheat}}\propto \Delta m_*/\Sigma_{\rm{gas}}$). The second assumption increases the gas consumption timescale in the inner disk.

In this paper, we will introduce a radial gas inflow prescription as a more realistic solution, i.e the gas from the outer disk flows inwards towards the inner disk. As we will now show, a plausible inflow prescription combined with a simpler star formation law
in which the star formation rate surface density is always proportional to the molecular gas surface density, produces gas surface density profiles that match observations without need for a radially-dependent SN feedback efficiency.

There are a number of physical mechanisms that can drive radial inflows of gas in the disk. The ones that are invoked most frequently involve gravitational interaction between gas in the disk and non-axisymmetric stellar structures such as bars and spiral structures (Kalnajs 1972). Simple physical considerations yield estimates of flow velocities ranging from 0.1 to a few km/s (Lacey \& Fall 1985; Bertin \& Lin 1996). Attempts to measure radial flow rates in galaxies have been confined to very small samples and have yielded inconclusive results (e.g. Wong, Blitz \& Bosma 2004; Haan et al. 2009; Zhang \& Buta 2012). Difficulties arise from the fact that flow patterns in individual galaxies are often irregular and disks are frequently not axisymmetric. In addition, different methods for estimating mass inflow rates yield discrepant results.

As a result, modellers generally resort to simple parameterized inflow prescriptions (e.g Lacey \& Fall 1985; Portinari \& Chiosi 2000; Sch\"{o}nrich \& Binney 2009; Spitoni \& Matteucci 2011). In this work, we have tested a number of inflow prescriptions. We first tried the simplest prescription in which the radial gas inflow velocity is constant for the whole disk (model (a) in Lacey \& Fall 1985), but we found that the inflow velocities required to move enough gas from the outer disk to the inner disk to compensate the depletion by star formation, lead to an unacceptably large pile-up of gas in the very inner disk. A prescription in which the rate of change of the angular momentum is proportional to the angular momentum yields results that agree best with observational data, i.e
\begin{equation}\label{eq:dl}
\frac{dL_{\rm{gas}}}{dt}=C {L_{{\rm{gas}}}}
\end{equation}
where $C$ is a constant for all galaxies. Because $L_{\rm gas}=m_{\rm gas}r_{\rm gas}v_{\rm cir}$, we have that
\begin{equation}\label{eq:vr}
v_{\rm inflow}=\alpha_v r
\end{equation}
The radial inflow velocity of the gas is thus proportional to the galactocentric radius $r$. This is equivalent to model (b) in Lacey \& Fall (1985). Spitoni \& Matteucci (2011) also invoke a radial flow prescription of this form to reproduce the observed stellar metallicity gradient in the Milky Way.

\begin{figure}
\centering
 \includegraphics[angle=0,scale=0.55]{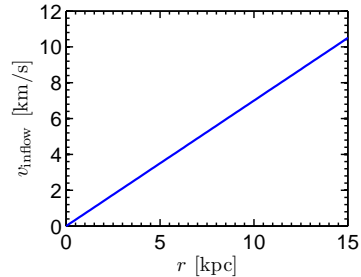}
 \caption
 {The relation between galactocentric radius $r$ and radial gas inflow velocity $v_{\rm inflow}$ from Eq. (\ref{eq:vr})
($\alpha_v=0.7~\rm{km~s^{-1}~kpc^{-1}}$ is adopted).
 }\label{fig:vinflow}
\end{figure}

The constant $\alpha_v$ in Eq. (\ref{eq:vr}) is an adjustable model parameter. As discussed later, $\alpha_v=0.7~\rm{km~s^{-1}~kpc^{-1}}$ is required to fit observations. This results in negligible flow velocities in the inner disk and flow speeds of several $\rm{km~s^{-1}}$ in the outer disk (see Fig. \ref{fig:vinflow}).

In the models, we calculate the gas inflow velocity according to Eq. (\ref{eq:vr}) after processing gas cooling and infall in each timestep and derive the new radius $r'=r-v_{\rm inflow} \Delta t$ for the gas in each ring, where $\Delta t$ is the length of the timestep. Proceeding from the inner ring to the outer ring, we move the gas to its new radius; metals associated with the cold gas move inward together with the gas.

\begin{figure*}
\centering
  \includegraphics[angle=-90,scale=0.45]{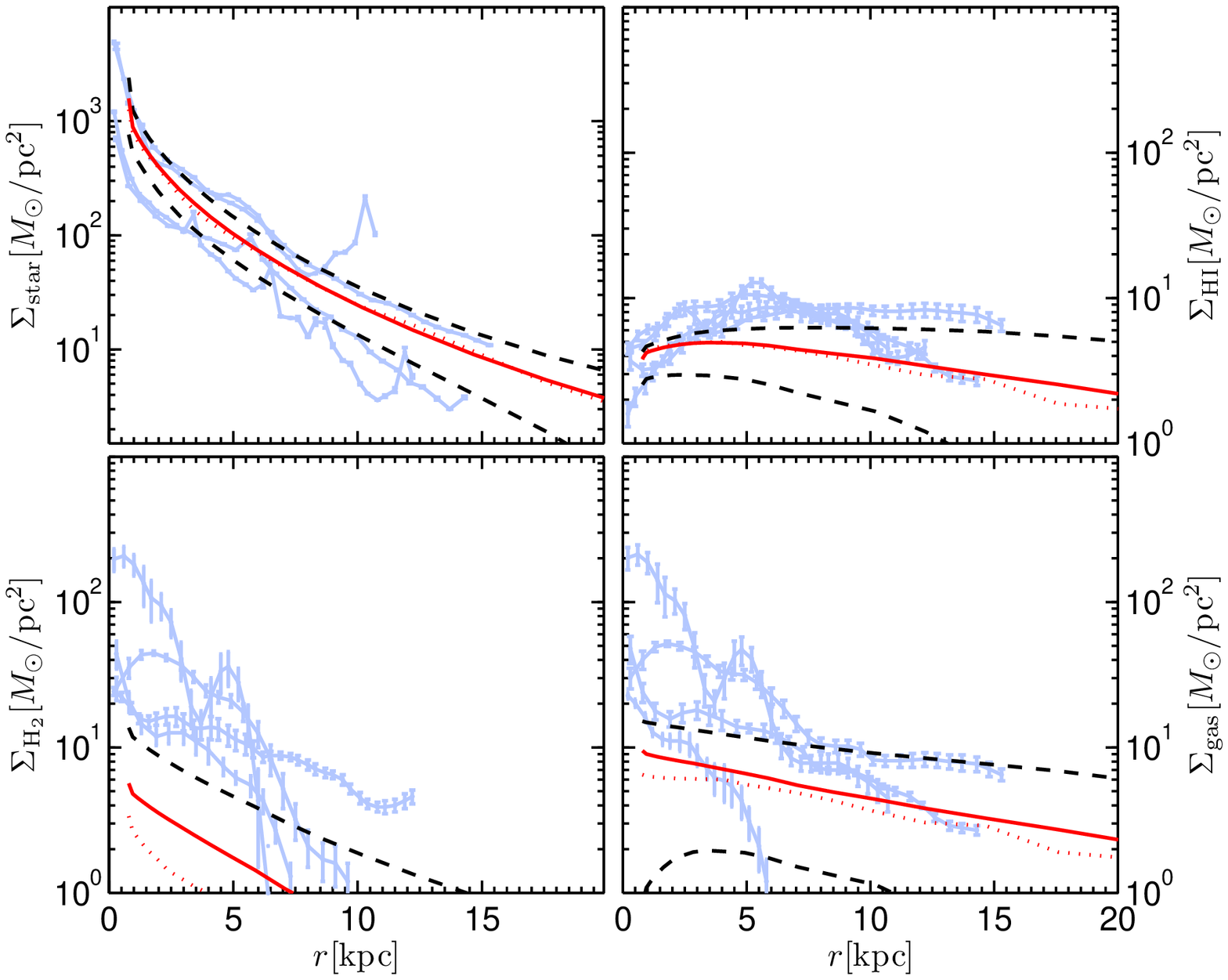}
  \includegraphics[angle=-90,scale=0.45]{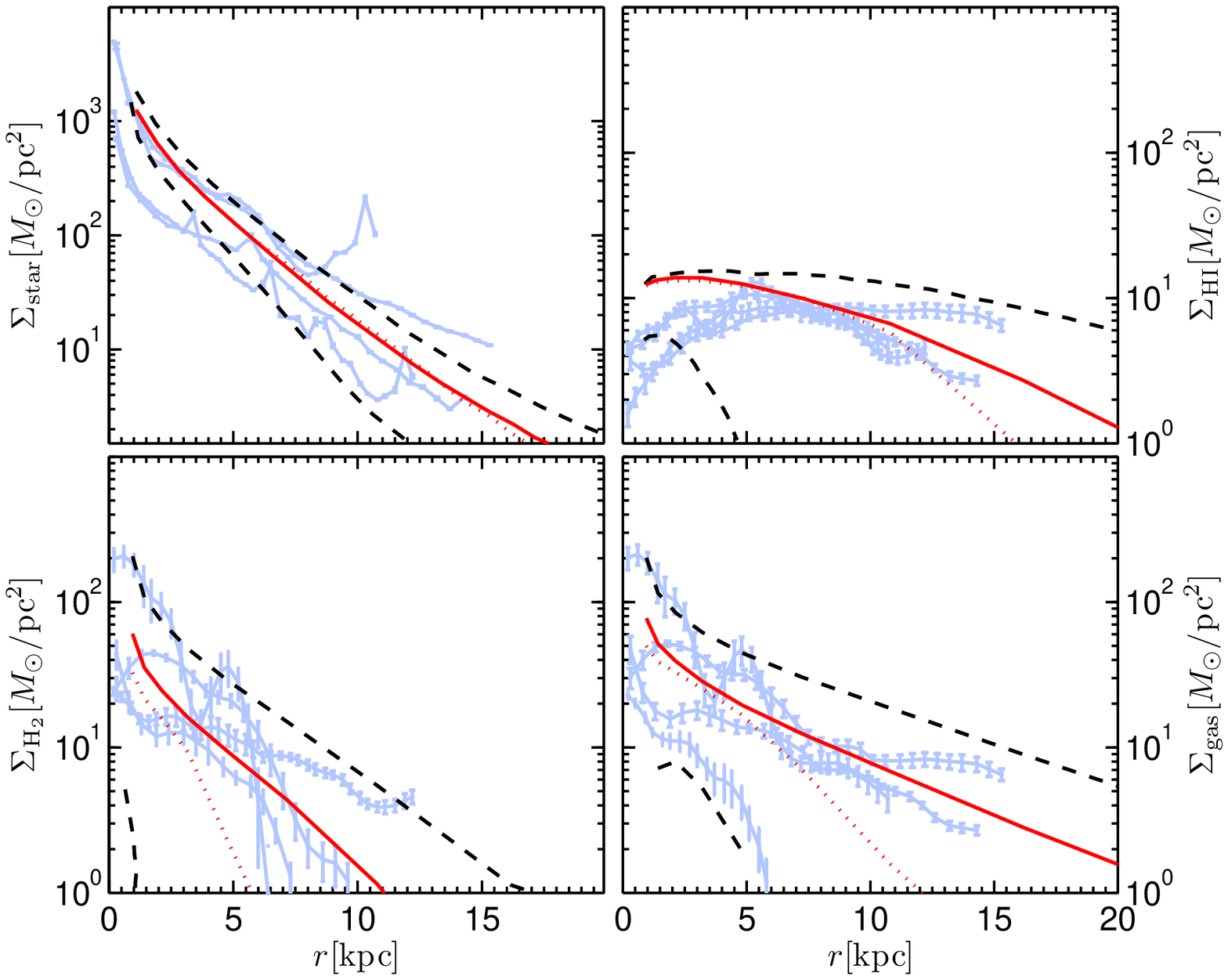}
  \caption
  {The radial surface density profiles of stars, atomic gas, molecular gas, and total cold gas for disk galaxies similar to the Milky Way. Results from the model with $\h2$ fraction prescription 1 (see Sec. \ref{chap:h2fraction}) are shown at at $z=0$. The two panels show the model results with (right panel) and without (left panel) radial gas inflow. The light blue curves with error bars are taken from the data of Leroy et al. (2008) for disk galaxies in the range of circular velocities $200 {\rm km s^{-1}}<v_{\rm cir}<235{\rm km s^{-1}}$ (NGC 0628, NGC 3184, NGC 5194, NGC 3521). The red solid and dotted curves are the mean and the median values from model results, and the black dashed curves are the $\pm1\sigma$ deviations about the mean values.
 }\label{fig:radialprofiles}
\end{figure*}

\begin{figure*}
\centering
  \includegraphics[angle=-90,scale=0.65]{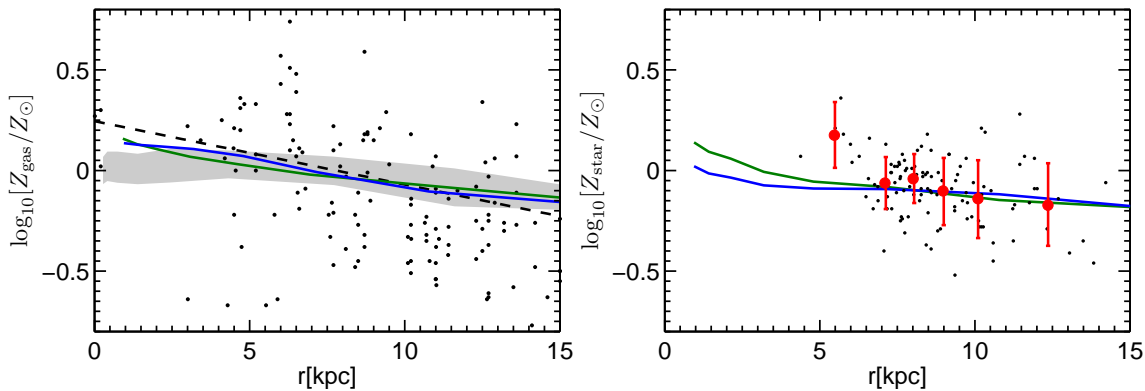}
  \caption
   {The mean radial metallicity profile of gas (left panel) and stars (right panel) for a Milky Way-sized galaxy at $z=0$ from models with (green curve) and without (blue curve) radial gas inflow. In the left panel, the black dots are the oxygen metallicity from HII regions in the Milky Way (Rudolph et al. 2006) and black dashed line is the corresponding linear fit to the data between 0 and 15 kpc. The gray area shows the $\pm 1\sigma$ deviation around the median radial profile of oxygen metallicity for galaxies with $10.5<\log_{10}[M_*/\ms]<11.0$ from Moran12. In the right panel, the black dots represent the stellar oxygen gradients from Cepheids (Andrievsky et al. 2002a,b,c;2004; Luck et al. 2003) and the red dots with error bars are corresponding mean values.
  }\label{fig:inflowz}
\end{figure*}

In Fig. \ref{fig:radialprofiles}, we show the best-fit radial surface density profiles of stars, atomic gas, molecular gas, and total cold (atomic+molecular) gas for disk galaxies in haloes with circular velocities comparable to that of the Milky Way ($200 < v_{\rm circ}< 235$ km/s) at $z=0$. The four panels on the left show the results if radial gas inflows are not included, and the four right panels show the results of our inflow model. Without inflows, both the molecular gas and average total cold gas radial surface density profiles are
too flat in comparison to the data of Leroy et al. (2008). The right panels show that the inclusion of radial flows leads to gas profiles that are in much better agreement with observations. We note that the stellar mass profiles are insensitive to the radial flow prescription; this is because the majority of the mass in the inner galaxy forms at high redshifts from gas cooling in smaller, denser haloes. The gas radial profiles are set by gas that cools at late times. Because the timescale over which $\h2$ is transformed into stars is $\sim 2$ Gyr (see Sec. \ref{chap:sfr}), the $\h2$ surface densities in the inner disk quickly drop to low values as a result of star formation and SN feedback, unless gas is transferred from the outer disk to the inner regions of the galaxy.

We note that we have tuned the parameters of the radial inflow model (in particular the constant $\alpha_v$ in Eq. \ref{eq:vr}) so that the gas radial profiles in Milky Way-type disks are reproduced. One might ask whether metallicity profiles provide an additional check on the inflow model. In Fig. \ref{fig:inflowz}, we show the mean radial stellar metallicity profile (right panel) and gas-phase metallicity profile (left panel) for the same set of Milky Way sized galaxies at $z=0$ (disk galaxies with $200 < v_{\rm circ}< 235$ km/s). Green curves show results with radial inflow and blue curves show models without radial gas inflow. In the right panel, the black dots represent the stellar oxygen measurements gradients derived from Cepheids by Andrievsky et al. (2002a,b,c;2004) and Luck et al. (2003). Red dots with error bars are the corresponding mean values. In the left panel, the black dots represent oxygen metallicity measurements from HII regions in the Milky Way by Rudolph et al. (2006) and the black dashed line is the corresponding linear fit to the data between 0 and 15 kpc, which exhibits a -0.031 dex/kpc for gas-phase metallicity gradient between 0 and 15 kpc. The metallicity gradient of the galaxies in our models is a bit shallower than found for the data (-0.018 dex/kpc for models without radial gas inflow and -0.02 dex/kpc for models with inflow), because the parameters of the models were tuned to fit the average gas-phase metallicity profiles of Milky Way mass galaxies from Moran12, which are flatter than the Milky Way disk (see Sec. \ref{chap:gasmetallicity} for detail). The gray area in the left panel shows the $\pm 1\sigma$ deviation around the median radial profiles of oxygen metallicity for galaxies with $10.5<\log_{10}[M_*/\ms]<11.0$ from Moran12.

We see from Fig. \ref{fig:inflowz} that inclusion of radial gas inflow yields gas and stellar metallicity gradients that are slightly steeper in the region of the disk interior to 3 kpc, but otherwise the predicted metallicity gradients are virtually indistinguishable in the two cases. In our model, infalling gas from the halo has already been pre-enriched by the ejection of metals during the early formation phase of the galaxy, and the outer disk metallicity is mainly affected by the metallicity of the gas accreted recently because of inside-out disk growth. On the other hand, very small inflow velocities in the inner disk only slightly change the inner disk metallicities. These two aspects mean that the radial gas inflow has only slight influence on the radial metallicity gradients. As we will discuss in Sec. \ref{chap:metalmixing}, the fraction of the metals produced by SNe we choose to put into the hot gas rather than the cold interstellar medium has a much stronger influence on metallicity gradients than the radial gas inflow.

\subsection{Prescriptions for the transition between atomic and molecular gas in the ISM} \label{chap:h2fraction}

In Fu10, we implemented two prescriptions for the transition between atomic and molecular gas in the interstellar medium. One is from Krumholz et al. (2009) ($\h2$ fraction prescription 1), who calculate an equilibrium $\h2$ fraction ($\fh2=\Sigma_{\h2}/(\Sigma_{\h2}+\Sigma_{\rm HI})$) for a spherical cloud with given dust content surrounded by a photo-dissociating UV field. In this prescription, the $\h2$ fraction is primarily a function of local cold gas surface density and metallicity. The second prescription ($\h2$ fraction prescription 2) originates from Elmegreen (1989, 1993) and Blitz \& Rosolowsky (2006), in which the $\fh2$ is a function of the pressure in the ISM. We adopt the approximation by Obreschkow \& Rawlings (2009), in which the local ISM pressure is expressed as a function of gas and stellar surface density in the radial ring. More details can be found in Section 3.2 of Fu10.

In this paper, we make a number of adaptations to the Krumholz et al. prescription. We update the prescription using the recent fitting equations in McKee \& Krumholz (2010) with the molecular gas fraction $\fh2$ given by
\begin{equation}\label{eq:mckeefh2}
\fh2=1-\frac{3}{4}\frac{s}{1 + 0.25s}
\end{equation}
for $s<2$ and $\fh2=0$ for $s\ge 2$. The $s$ in Eq. (\ref{eq:mckeefh2}) is defined as
\begin{equation}\label{eq:mckees}
s = \frac{{\ln \left( {1 + 0.6\chi + 0.01{\chi ^2}} \right)}}{0.6{\tau_c}}
\end{equation}
in which $\chi=3.1\left(1+3.1Z'^{0.365}\right)/4.1$ and $\tau_c =0.66\left(\Sigma_{\rm comp}/\ms{\rm{pc}}^{-2}\right)Z'$. Note that $Z'=Z_{\rm gas}/Z_{\odot}$ is the gas-phase metallicity relative to the solar value, and $\Sigma_{\rm comp}$ is the gas surface density of the giant gas cloud. Since the gas surface density in the model is the azimuthally averaged value in each concentric ring, a clumping factor $c_{\rm f}$ is introduced to consider the difference between $\Sigma_{\rm comp}$ and $\sgas$
\begin{equation}\label{eq:cf}
\Sigma_{\rm comp}=c_{\rm f}\sgas
\end{equation}

One problem with the Krumholz et al. prescription is that it can easily yield non-convergent results. The reason for this is that the value of $f_{\h2}$ is very sensitive to the exact value of the gas-phase metallicity when the metallicity is well below the solar value (see Figure 5 in McKee \& Krumholz 2010). Molecular clouds do not form until metals have been produced, and metal production is dependent on the formation of stars, which can only take place in molecular clouds, so the conditions for star formation to occur in the first generation of haloes to form in the simulation are subject to considerable uncertainty. Moreover, at the resolution limit of the simulation, the assembly histories of haloes, and hence the star formation histories of galaxies embedded within them, cannot be tracked accurately. It thus becomes very difficult to get results that are consistent for galaxies forming in haloes of the same mass in simulations of different resolutions (e.g. MS and MS-II). Similar problems were reported by Kuhlen et al. (2012) in their attempts to model dwarf galaxies with Krumholz et al. prescription.

\begin{figure*}
\centering
 \includegraphics[angle=-90,scale=0.6]{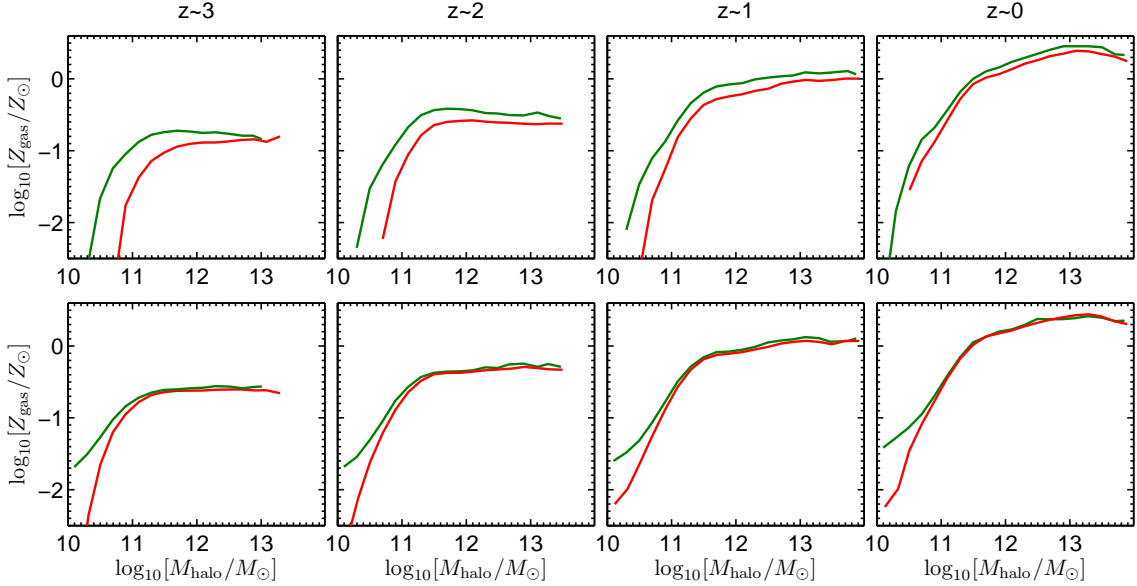}
 \caption
 {The relation between gas-phase metallicity and galaxy halo mass at redshifts 3, 2, 1, and 0. The red curves and green curves are the mean values derived for the MS and the MS-II simulations, respectively. The four top panels are from the Krumholz et al. $\h2$ prescription
 with fixed clumping factor $c_{\rm f}=1.5$, and the four bottom panels are from the Krumholz et al. $\h2$ prescription with variable clumping factor dependent on gas-phase metallicity (see text).
 }\label{fig:zgasmhalo}
\end{figure*}

In the top panels of Fig. \ref{fig:zgasmhalo}, we plot the relation between the gas-phase metallicity of a central galaxy and the mass of its parent halo at different redshifts. The red curves and green curves show results for the MS and MS-II simulations, respectively. We have adopted $c_{\rm f}=1.5$ for all galaxies at all redshifts, as in Fu10. As can be seen, at fixed halo mass, the gas-phase metallicity is higher in the MS-II run than in the MS run, particularly for low mass haloes at high redshifts. Since the halo resolution of MS-II is 125 times higher than MS, a larger fraction of the formation and chemical enrichment history of low mass haloes lies below the resolution limit for MS than MS-II. Because the Krumholz et al. prescription is so sensitive at low metallicities, the resolution discrepancies are enhanced for high redshift galaxies in low mass haloes.

The question arises as to why stars are able to form at all in low-metallicity galaxies. For example, 1ZW18 is the most metal-poor galaxy known, yet it is a starburst system (e.g. Martin 1996). The main reason is that the gas in low-metallicity dwarf galaxies is clumpy on scales of $\sim$100 pc (Lo, Sargent \& Young 1993; Stil \& Israel 2002). To account for this, Kuhlen et al. (2012) adopt a clumping factor $c_{\rm f}=30$ for their models of dwarf galaxies at high redshift; Luo et al. (2011) adopt $c_{\rm f}=20$ to model high-redshift damped Lyman-alpha systems.

In this paper, we adopt a variable clumping factor that depends on gas-phase metallicity. We use the following equation to describe the relation between gas-phase metallicity and clumping factor
\begin{equation}\label{eq:cfz}
c_{\rm f}=Z'^{-0.7}
\end{equation}
for $0.01<Z'<1$ and $c_{\rm f}=1$ for $Z'\ge 1$. The clumping factor for $\log_{10}Z' > -1$ is between 1 and 5, which agrees with the values suggested for normal galaxies in KMT09.

In the four bottom panels of Fig. \ref{fig:zgasmhalo}, we plot central galaxy gas-phase metallicity as a function of halo mass for models that adopt a variable $c_{\rm f}$. Interestingly, the difference between MS-I and MS-II results is largely eliminated for galaxies in haloes with larger than $10^{11}\ms$, because the higher clumping factor compensates the low molecular gas surface densities in dwarf galaxies and high redshift galaxies with low gas-phase metallicities. Of course, this does not mean that our clumping factor prescription is correct! It just means that we are able to match together the results from the two simulations with greater ease.

\subsection{Star formation prescription} \label{chap:sfr}

DLB07 and Guo11 both adopt the following star formation law which relates the star formation rate surface density to the cold gas surface
density of the ISM, $\dot m_* = \alpha \left( m_{\rm gas}-m_{\rm crit}\right)/t_{\rm dyn}$ (see Eq. (\ref{eq:sfr0}) in Appendix). In Fu10, we adopt a two-regime star formation law related to both molecular and atomic gas phases
\begin{equation}\label{eq:sfr}
\Sigma_{\rm{SFR}} =
\begin{cases}
   \alpha\Sigma_{\rm{H_2}} & (f_{\rm{H_2}}\ge 0.5)  \\
   \alpha'\Sigma_{\rm{gas}}^2 & (f_{\rm{H_2}} < 0.5)  \\
\end{cases}
\end{equation}
The star formation law in atomic gas dominant regions helps solve the problem of gas consumption times being too slow in outer disks.

Since we now include radial gas inflow process in Sec. \ref{chap:gasflow} to solve the problem of gas consumption timescales in both inner and outer disks, we now abandon the complex two-regime star formation form Eq. (\ref{eq:sfr}) in the model. We adopt a simple law in which the star formation surface density is proportional to the $\h2$ surface density: $\Sigma_{\rm SFR}\propto\Sh2$ (e.g Leroy et al. 2008; Bigiel, Leroy \& Walter 2011; Schruba et al. 2011; Leroy et al. 2013). We adopt a constant molecular gas consumption timescale for all galaxies at all redshifts. There is evidence that molecular gas consumption times may be longer in massive, early-type galaxies than in normal, present-day star-forming spirals (Saintonge et al. 2011b) and shorter in starburst galaxies (Genzel et al. 2010), but we will neglect these complications, because it is still unclear what the underlying physical cause of these effects are.

The star formation law adopted in the model can be simply written as
\begin{equation}\label{eq:h2sfr}
\Sigma_{\rm SFR}=\alpha_{\h2} \Sigma_{\h2}
\end{equation}
in which the $\h2$ star formation efficiency $\alpha_{\h2}=5.3\times10^{-10}\rm{yr^{-1}}$ is slightly adapted to yield a better fit to the results stellar mass function at $z=0$. We note that the molecular gas surface density $\Sigma_{\h2}$ in Eq. (\ref{eq:h2sfr}) does not include the helium component.

\subsection{The gas-phase radial abundance gradients and the mixing of metals} \label{chap:metalmixing}

In the models of DLB07, Guo11 and Fu10, metals produced by star formation in the disk are mixed instantaneously with the cold gas in the disk (see the sixth paragraph in Appendix). SN reheating results in part of the cold gas (along with the metals that have been mixed into it) being transferred to the hot gas in the halo. If the SN explosion energy $E_{\rm SN}$ is larger than the energy required to reheat the cold gas $E_{\rm reheated}$, the remaining energy $E_{\rm SN}-E_{\rm reheated}$ will eject part of the halo hot gas out of the halo: this corresponds to the so-called ``ejected component'' (see Fig. \ref{fig:processes}, and Eq. (\ref{eq:snenergy}), (\ref{eq:reheat}), (\ref{eq:meject})).

In the models, the gas-phase metallicity radial gradient is mainly determined by two processes. One is the fraction of metals from star formation and SN explosions that remain in the interstellar cold gas, which mainly determines the gas-phase metallicity in the inner disk. The other one is the metallicity of recent infalling gas pre-enriched by the SN ejecta in early epochs, which mainly determines the gas-phase metallicity in outer disk because of the inside-out disk growth (e.g Pilkington et al. 2012). This is why the fraction of metals produced by young stars that is injected into the hot gaseous halo can affect the slope of the metallicity gradient.

On the other hand, we note that there is direct observational evidence from deep X-ray spectral imaging of nearby star-forming disk galaxies that the hot gas produced by supernovae type II (SNe-II) is highly metal-enriched (Martin, Kobulnicky \& Heckman 2002) and also enhanced in alpha elements (Strickland et al. 2004). Even in ``normal'' spirals, the gas is directly observed at scale heights 4-8 kpc above the galactic disk, before its emissivity decreases below the detection threshold (Strickland et al. 2004). The scenario favoured by current observations is that SN feedback in the disks of star-forming galaxies create exponential atmospheres of hot gas via blow-out and venting of hot gas from the disk.

\begin{figure}
\centering
  \includegraphics[angle=-90,scale=0.50]{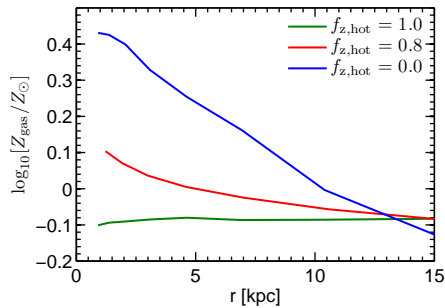}\\
  \caption{The radial gas-phase metallicity gradient for disk galaxies with stellar mass $10^{10.0}<M_*/M_{\odot}<10^{10.5}$ from the model results at $z=0$. The 3 different colour curves represent the model results with $f_{\rm z,hot}=0.0, 0.8, 1.0$, respectively.}\label{fig:FracZtoHot}
\end{figure}

We have experimented with mixing a fraction $f_{\rm z,hot}$ of the metals produced by star formation directly with the hot gas in the halo, i.e a fraction of the metal from the star is directly recycled to the hot gas in Fig. \ref{fig:processes}. This turns out to have very strong impact on the predicted gas-phase metallicity gradients in the galaxy. In Fig. \ref{fig:FracZtoHot}, we plot the mean radial gas-phase metallicity gradients for disk galaxies with stellar mass $10^{10.0}<M_*/M_{\odot}<10^{10.5}$ at $z=0$. The blue curve shows results for $f_{\rm z,hot}=0.0$, which is effectively the DLB07 and Guo11 mixing prescription, in which none of the metals are directly mixed into the hot gas. This yields a steep gradient, with more than 0.6 dex change in metallicity in the inner 15 kpc of the galaxy. The green curve shows what happens if we adopt $f_{\rm z,hot}=1.0$, i.e. all metals are mixed with the hot gas. In this case, the metallicity gradient nearly disappears, because we assume the hot gas in the halo is fully mixed and the infalling gas onto the disk has a uniform metallicity at different radii. In this paper, we assume that metals from AGB star winds are retained in the interstellar cold gas and metals from core-collapse SN explosions are mixed into hot gaseous halo. According to the yield data compiled by Marigo et al. (2001) for AGB stars and Portinari, Chiosi \& Bressan (1998) for SNe-II, about 20 percent of the total metal yield comes from AGB stars of low and intermediate masses when we adopt a Kroupa IMF between 0.8$\ms$ and 120 $\ms$. \footnote {In our current model, we adopt instantaneous recycling approximation for the metal production and chemical evolution, and we only consider the metal yield from AGB stars and SNe-II. The mass of metal elements ejected from AGB stars is about $0.1\ms$ per star and about $10\ms$ per star from SNe-II (see Figure 2 and 4 in Yates et al. 2013 for the yield data from Marigo 2001 and Portinari et al. 1998). We adopt a Kroupa IMF and assume that stars with masses in the range $0.8<M/\ms<8$ end their lives as AGB stars, and that stars with masses in the range $8<M/\ms<120$ explode as SNe-II. This results in an estimate of $20\%$ of the metal yield from AGB stars and $80\%$ from SNe-II. Our future work will include more detailed and accurate prescriptions for chemical evolution, similar to Yates et al. (2013).} The red curve in Fig. \ref{fig:FracZtoHot} shows the case where metals from AGB star outflows are retained in the cold ISM and metals from core-collapse SN explosions are mixed into hot gas ($f_{\rm z,hot}=1-0.2=0.8$).

In Sec. \ref{chap:gasmetallicity}, we will compare the model gas-phase metallicity gradients to observations and also explore how the predicted metallicity profiles depend on galaxy properties such as mass and gas mass fraction. We will show that $f_{\rm z,hot}=0.8$ gives reasonably good agreement with observations.

\subsection{The parameters in the models} \label{chap:parameter}

As discussed in Fu10, we tune some of the model parameters to fit the observed stellar, HI and $\h2$ mass functions at $z=0$. In the top part of Tab. \ref{tab:parameters}, we list the model parameters which are changed in this paper with respect to those listed in previous papers. The value of $\h2$ star formation efficiency $\alpha_{\h2}$, the amplitude of SN reheating efficiency $\epsilon$, and the quiescent hot gas black hole accretion rate $\kappa_{\rm AGN}$ (see Eq. (\ref{eq:radio})) are tuned to fit the amplitudes of both the stellar and the gas mass functions. The normalization of the radial gas inflow rate $\alpha_v$ in Sec. \ref{chap:gasflow} is tuned to reproduce the gas surface density profiles of Milky Way-type galaxies and to fit the observed HI and $\h2$ mass functions. The fraction of metals directly mixed with halo hot gas is chosen to be $0.8$. Other parameters are not changed in previous papers (bottom section of Tab. \ref{tab:parameters}): $f_{\rm b}$, $f_{\rm BH}$, $T_{\rm merger}$, $Y$ are the same as in Table 1 in Croton et al. (2006) (they remain unchanged in DLB07, Fu10, Guo11 and Guo13); $R$ is same as in DLB07 (it remains unchanged in Guo11 \& Guo13); $\beta_1$, $V_{\rm reheat}$, $\eta$, $\beta_2$, $V_{\rm eject}$ are same as in Table 2 in Guo13; $\gamma$ is the same as in Table 1 in Guo11 (it remains unchanged in Guo13); $\xi$, $P_0$, $\alpha_P$ are the same as in Table 1 of Fu10.

\begin{table*}
 \centering
 \caption{The model parameters introduced/changed in this paper (top section) and other parameters whose values are the same as
in previous papers (bottom section). Note that the parameters marked as ``Croton et al. (2006)'' remain unchanged in DLB07, Fu10, Guo11 and Guo13.\label{tab:parameters}}
 \begin{tabular}{|c|l|l|l|}
 \hline \hline
Parameter & Value & Description & Remark\\
 \hline
$\alpha_v$      & $0.7~\rm{km~s^{-1}~kpc^{-1}}$ & the ratio of radial gas inflow and gas radius & Eq. (\ref{eq:vr})\\
$\alpha_{\h2}$ & $5.3\times10^{-10}\rm{yr^{-1}}$ & $\h2$ star formation efficiency          &  Eq. (\ref{eq:h2sfr}) \\
$\epsilon$     & 5.0 & Amplitude of SN reheating efficiency     & Tab. 1 in Guo11 \& Tab. 2 in Guo13\\
$\kappa_{\rm{AGN}}$ &$1.5\times{10^{-5}}\ms\rm{yr^{-1}}$ & Quiescent hot gas black hole accretion rate & Eq. (\ref{eq:radio})\\
$f_{\rm z,hot}$ & 0.8 & Fraction of metal elements from quiescent star & Sec. \ref{chap:metalmixing} \\
                &     & formation directly mixed with halo hot gas     & \\
\hline
$f_{\rm b}$&0.17& Cosmic baryon fraction & Croton et al. (2006)\\
$f_{\rm BH}$& 0.03 & Merger cold gas BH accretion fraction & Croton et al. (2006)\\
$T_{\rm merger}$ & 0.3 & Major merger mass ratio threshold & Croton et al. (2006)\\
$Y$ & 0.03 & Yield of metals produced per unit star formation & Croton et al. (2006)\\
$R$ & 0.43 & Instantaneous recycled fraction of star  & DLB07, Guo11 \& Guo13\\
         && formation to the cold gas &\\
$\beta_1$& 3.2 & Slope of SN reheating efficiency & Guo13\\
$V_{\rm reheat}$ & $80~\rm{km~s^{-1}}$ & Normalization of SN reheating efficiency & Guo13\\
$\eta$ & 0.18 & Amplitude of SN ejection efficiency & Guo13\\
$\beta_2$ &3.2 & Slope of SN ejection efficiency & Guo13\\
$V_{\rm eject}$ & $90~\rm{km~s^{-1}}$ & Normalization of SN ejection efficiency & Guo13\\
$\gamma$ & 0.3 & Ejecta reincorporation efficiency & Guo11 \& Guo13\\
$\xi$ & 1.3 & Warm-phase correction factor & Fu10\\
$P_0$, $\alpha_P$ & $5.93\times10^{-13}$ Pa, 0.92 & Constant and index of the relation   & Fu10\\
 &   & between molecular ratio and ISM pressure & \\
\hline \hline
\end{tabular}
\end{table*}

\section{Stellar and gas mass functions at $\boldsymbol{z=0}$} \label{chap:massfunction}

The Fu10 models use the SN feedback prescriptions in DLB07, resulting in stellar, HI and $\h2$ mass functions that are too steep at low masses (see Figure 1 in Fu et al. 2012). In this paper, we use the prescriptions in Guo11, which provide a better fit to the low-mass end of the stellar mass function.

\begin{figure*}
\centering
  \includegraphics[angle=-90,scale=0.6]{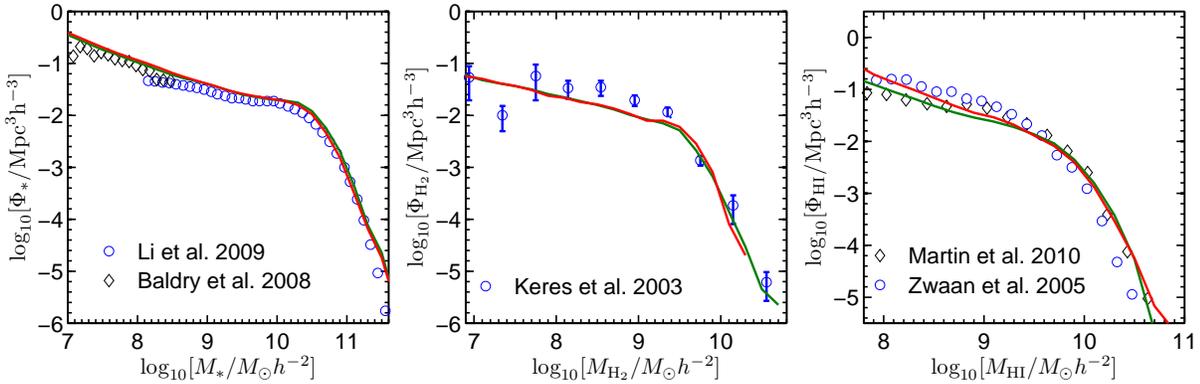}\\
  \caption{The stellar, $\h2$ and HI mass functions from model galaxies at $z=0$ compared with the observations. The red solid curves and green solid curves in each panel are from the models with $\h2$ fraction prescription 1 and 2, respectively. The observed stellar mass functions are from Li \& White (2009) and from Baldry et al. (2008). The observed $\h2$ mass function is from the FCRAO CO survey by Keres et al. (2003). The observed HI mass function is from Zwaan et al. (2005) using HIPASS data, and Martin et al. (2010) using ALFALFA data.
  }\label{fig:massfunction0}
\end{figure*}

In Fig. \ref{fig:massfunction0}, we show the mass functions of stars, molecular gas and atomic gas for our revised models. Red solid curves show results for $\h2$ fraction prescription 1 and green solid curves for $\h2$ fraction prescription 2. We have computed the mass functions using the combined results from running the code on the MS and MS-II simulations. Blue circles show the stellar mass function derived from the Data Release 7 of SDSS by Li \& White (2009); black diamonds show SDSS Data Release 4 results from Baldry et al. (2008). The observed $\h2$ mass function was derived from the FCRAO survey (Young et al. 1995) assuming a constant CO-$\h2$ conversion factor by Keres, Yun \& Young (2003). The HI mass functions are from Zwaan et al. (2005) (blue circles) and Martin et al. (2010) (black diamonds).

As mentioned in Sec. \ref{chap:parameter}, we tune the model parameters to fit the HI, $\h2$ and stellar mass functions, so it is no surprise that the agreement with models is very good.

\section{Star formation gradients} \label{chap:sfrsize}

In this section, we compare the star formation rate gradients in our model galaxies with observations. This provides an important test of whether the present-day growth of disks predicted by the models agrees with the data.

We have adopted two observational samples for the analysis. Leroy et al. (2008) analyzed star formation gradients in a sample of 23 nearby spiral galaxies from the THINGS/HERACLES survey, finding that the scale length of the star formation rate surface density profile $l_{\rm SFR}$ is roughly proportional to the scale length of the stellar disk $l_*$ ($l_{\rm SFR}=(1\pm0.2)l_*$). We also make use of star formation rate surface density measurements from the MMT long-slit observations of 174 galaxies from Moran12. 119 of these galaxies (i.e. around $70\%$ of the total sample) have emission lines strong enough to measure both SFR and gas-phase metallicity gradients out to 2$r_{50}$.

\begin{figure*}
\centering
  \includegraphics[angle=-90,scale=0.6]{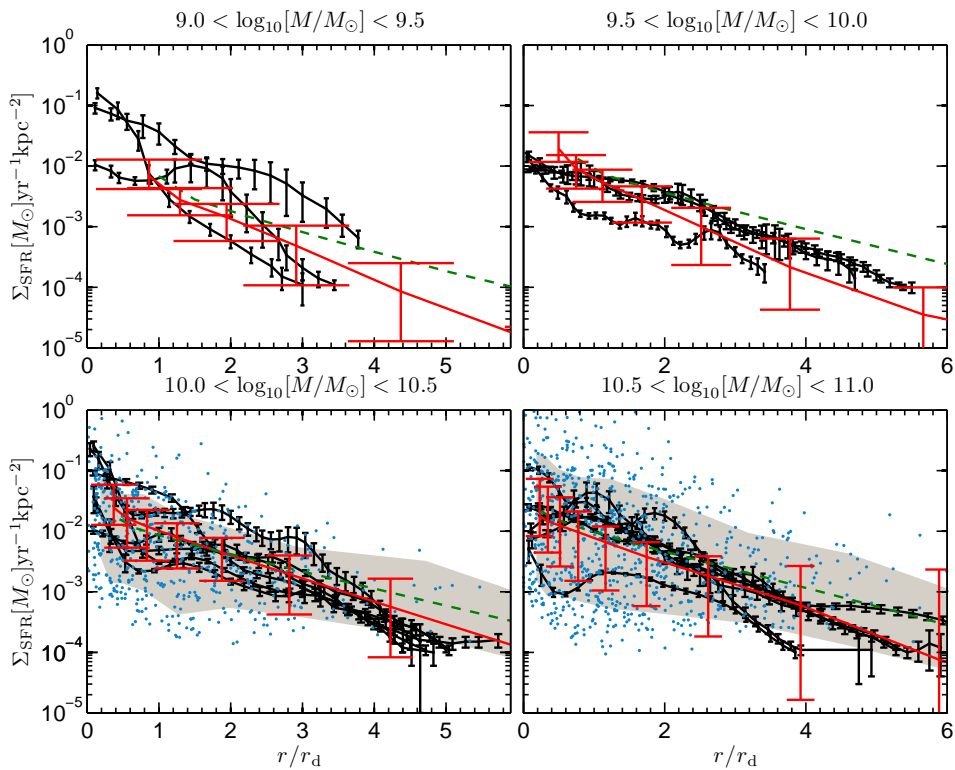}\\
  \caption{The radial profiles of star formation surface density for model galaxies at $z=0$ compared with observations from Leroy et al. (2008) (black curves with error bars) and Moran12 (blue dots). The gray areas represent the $\pm1\sigma$ deviations around the median values for the Moran12 data. The red solid curves show model results of mean radial profiles for $\h2$ prescription 1 and the green dashed curves are for $\h2$ fraction prescription 2. The error bars on the red solid curves represent $\pm1\sigma$ scatter about the mean values for the model. The panels show results in 4 stellar mass bins: $10^{9.0}<M_*/M_{\odot}<10^{9.5}$, $10^{9.5}<M_*/M_{\odot}<10^{10.0}$, $10^{10.0}<M_*/M_{\odot}<10^{10.5}$, $10^{10.5}<M_*/M_{\odot}<10^{11.0}$. The radii plotted on the x-axis are scaled by dividing by the disk scale length $r_d$.
 }\label{fig:sfr1}
\end{figure*}

In Fig. \ref{fig:sfr1}, we plot the star formation rate surface density as a function of scaled radius for model galaxies in four different stellar mass bins: $9.0<\log_{10}[M_*/\ms]<9.5$, $9.5<\log_{10}[M_*/\ms]<10.0$, $10.0<\log_{10}[M_*/\ms]<10.5$, and $10.5<\log_{10}[M_*/\ms]<11.0$. For the model sample, $r_d$ is calculated by fitting exponentials to the stellar profiles of the disk.
The red solid and green dashed curves show results of mean radial profiles for $\h2$ fraction prescriptions 1 and 2, respectively, and the error bars on the red solid curves represent the $\pm1\sigma$ scatter between different model galaxies.

In each panel, black curves with error bars are star formation surface density profiles from Leroy et al. (2008); $r_d$ for each galaxy is taken from Table 4 of that paper. The light blue dots are from Moran12, who obtained long-slit spectra of 174 star-forming galaxies with stellar masses greater than $10^{10} M_{\odot}$ from the GALEX Arecibo Sloan Digital Sky Survey (GASS) survey (Catinella et al. 2010). In this case, $r_d$ is estimated from $r_{90}$ assuming an exponential profile (i.e. $r_{90}=3.9r_d$). Note that we have simply plotted the star formation rate surface density measured for each spectral bin along the slit, i.e. these data points are not averaged in radial bins as for the Leroy et al. data, so the scatter from one galaxy to another will be substantially larger. The grey shaded region shows the $1\sigma$ scatter around the median for the Moran12 sample. As can be seen, the agreement between the Moran et al. and the Leroy et al. data is quite good.

Our results for $\h2$ fraction prescription 1 agree well with the data, particularly for the 3 highest stellar mass bins. $\h2$ fraction prescription 2 yields higher star formation rate surface densities in the outer disks. This is because the pressure-based prescription
yields higher $\h2$ fractions in regions of low gas surface density with low gas-phase metallicities.

We note that we tuned the radial inflow prescriptions to match the $\h2$ surface density profiles of the galaxies in the Leroy et al. sample for stellar masses $\log_{10}[M_*/\ms]\sim 10.6$. Our star formation prescription is also motivated by results obtained for this sample. This means that star formation rate surface density profiles for galaxies with $\log_{10}[M_*/\ms]\sim 10.6$ will match the
data, essentially by construction. The main check in this section is to test whether we can match to SFR profiles over a large range in stellar mass. One significant conclusion that we reach is that $\h2$ fraction prescription 1 does a better job than $\h2$ prescription 2 at matching the outer gas profiles of galaxies with stellar masses less than $10^{10} M_{\odot}$.

\section{Gas-phase metallicity gradients in galaxies} \label{chap:gasmetallicity}

In this section, we will compare the gas-phase metallicity gradients predicted by the model with data from Moran12. We will first study how the gradient depends on the stellar mass of the galaxy, and then examine its dependence on the global gas fraction. When we compare the model with data from Moran12, we choose galaxies that have HI and $\h2$ gas mass fractions greater than $\sim 1-3\%$, such that HI and CO lines would be detected in the GASS and COLD GASS surveys: i.e. $\log_{10}[\HIs]>-1.82$ for galaxies with $\log_{10}[M_*/\ms]>10.3$, and $\log_{10}[\HIs]>-1.066\log_{10}[M_*/\ms]+9.16$ for galaxies with $10.0<\log_{10}[M_*/\ms]<10.3$ (Kauffmann et al. 2012). This is a slightly more stringent cut than adopted by Moran12 for his emission-line analysis, but we have verified that the precise location of the cut makes negligible difference to all the results presented in this paper.

We adopt the gas-phase oxygen abundance based on the O3N2 empirical index described by Pettini \& Pagel (2004), which is derived from the
$\rm{N}\left[{\rm{II}} \right]\lambda 6583/{\rm{H}}\alpha$ and $\rm{O}\left[{\rm III}\right]\lambda 5007/{\rm{H}}\beta$ emission line ratios (see Section 3.1 of Moran12 for more details).

In the models, we only track total metallicity; the value of $Z_{\rm gas}/Z_{\odot}$ represents the cold gas-phase metallicity in units of the solar value. To make a comparison to oxygen abundance from observations, we adopt $12+\log_{10}(\rm{O/H})=8.69$ as the solar value (Asplund et al. 2009). We convert $Z_{\rm gas}$ to $(\rm{O/H})_{\rm gas}$ using the equation
\begin{equation}\label{eq:OZgas}
{12+\log_{10}(\rm{O/H})_{\rm gas}}=\log_{10}[Z_{\rm gas}/Z_{\odot}]+8.69
\end{equation}

\subsection{Gas-phase metallicity gradients as a function of stellar mass} \label{chap:dz-ms}

\begin{figure*}
\centering
  \includegraphics[angle=-90,scale=0.60]{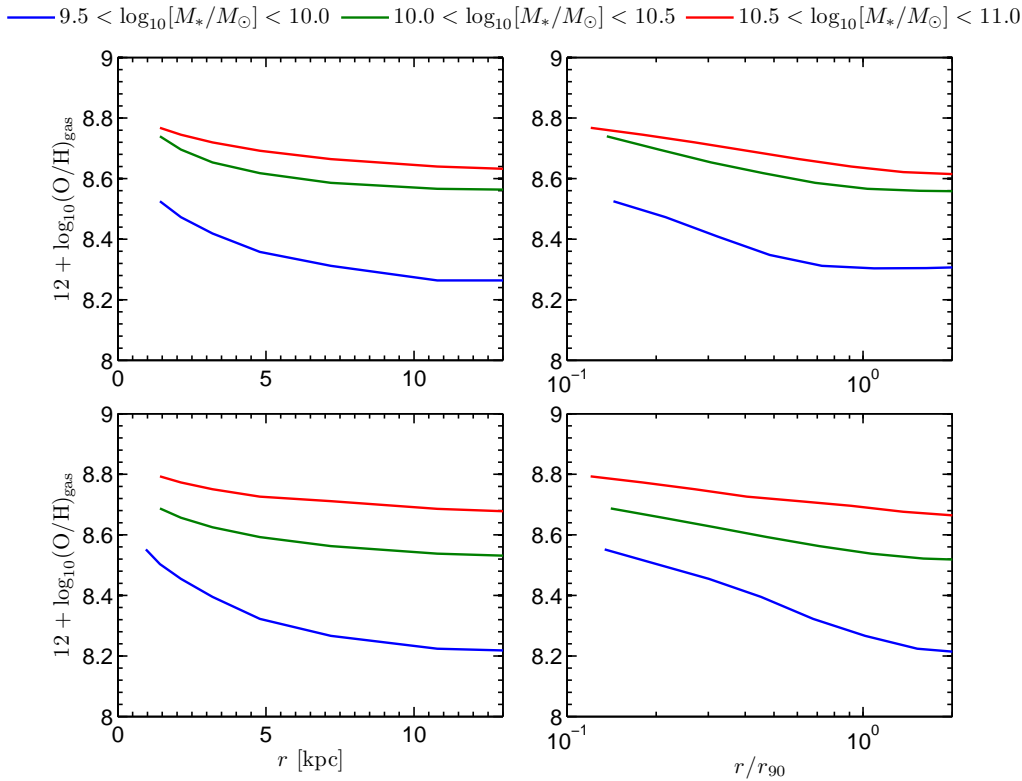}\\
  \caption{The mean radial profiles of gas-phase metallicity from model galaxies at $z=0$ in 3 mass bins: $10^{9.5}<M_*/M_{\odot}<10^{10.0}$ (blue), $10^{10.0}<M_*/M_{\odot}<10^{10.5}$ (green), $10^{10.5}<M_*/M_{\odot}<10^{11.0}$ (red). The top two panels are the results from $\h2$ fraction prescription 1, and bottom two panels are from $\h2$ fraction prescription 2. The radii plotted on the x-axis are scaled by divising by the disk scale length $r_d$.}\label{fig:zprofiles1}
\end{figure*}

In Fig. \ref{fig:zprofiles1}, we plot mean gas-phase metallicity profiles for model galaxies in 3 stellar mass bins: $10^{9.5}<M_*/M_{\odot}<10^{10.0}$, $10^{10.0}<M_*/M_{\odot}<10^{10.5}$, and $10^{10.5}<M_*/M_{\odot}<10^{11.0}$. The top two panels show results for $\h2$ fraction prescription 1 and the bottom two panels are for $\h2$ fraction prescription 2. In the left panels, metallicity is plotted as a function of radius in kpc, and in the right panels it is plotted as a function of radius scaled by $r_{90}$, the radius enclosing $90\%$ of the total stellar mass of the galaxy. The main result is that the gas radial metallicity profiles are flatter in higher mass galaxies; this is seen both in the left and in the right panels, where the radius has been scaled by the size of the galaxy.

We have investigated the {\em cause} of varying metallicity gradients in the models by examining how the metallicity gradient in a central galaxy of a given halo evolves over time. We have found that the evolution of the gas-phase metallicity gradient is most closely tied
to the {\em merger history} of the galaxy. The starbursts induced by the mergers consume all the cold gas (in the case of major mergers) or a large fraction of it (in the case of minor mergers) (see Sec. \ref{chap:bulgeprofiles}), which destroy the radial gas-phase metallicity gradient formed in early epochs. The metallicity gradient is re-established by the new gas accreted after mergers, and as a result, the gas-phase radial metallicity gradient in the new disk will be weaker. The models thus predict that gas-phase metallicity gradients correlate more strongly with the bulge mass fraction of the galaxy than with its stellar mass. The correlation with stellar mass arises because massive galaxies have experienced more mergers and have larger bulge mass fractions than less massive galaxies.

\begin{figure*}
\centering
  \includegraphics[angle=-90,scale=0.65]{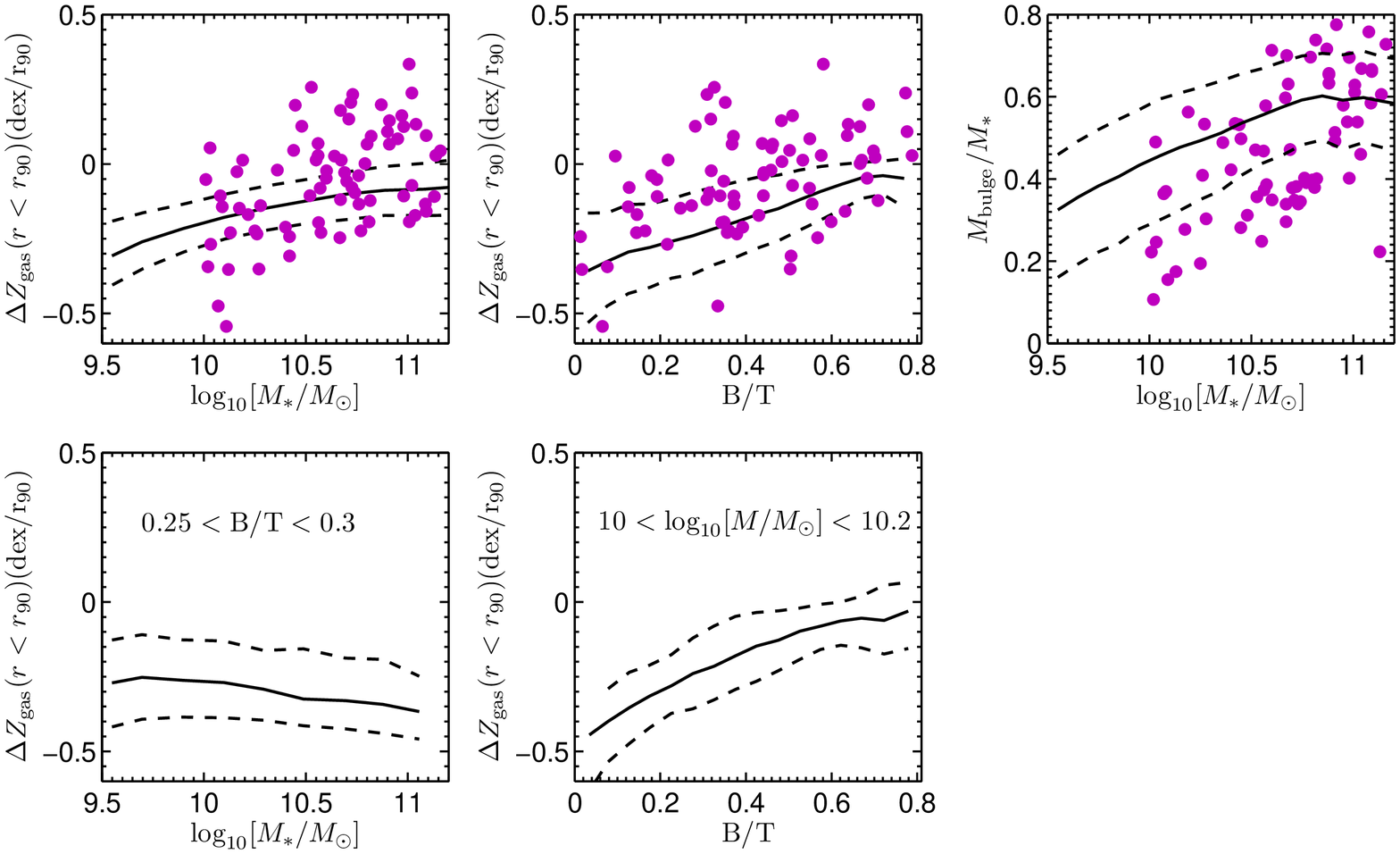}\\
  \caption{Top left panel: gas-phase metallicity (in units of dex/$r_{90}$) vs. stellar mass; top middle panel: gas-phase metallicity vs. bulge-to-total ($B/T$) ratio; top right panel: the $B/T$ ratio vs. stellar mass; bottom left panel: gas-phase metallicity vs. stellar mass for model galaxies with $B/T$ ratio in the range $0.25<B/T<0.3$; bottom right panel: gas-phase metallicity vs. $B/T$ ratio for model galaxies in the stellar mass range $10.0<\log_{10}[M_*/\ms]<10.2$. In each panel, the black solid curves show the mean values from the model sample. The black dashed curves indicate the $\pm1\sigma$ scatter around the mean. The purple dots show the results of individual spectral bin measurements from the Moran12 data set.}
  \label{fig:mstarzgas}
\end{figure*}

This is illustrated in detail in Fig. \ref{fig:mstarzgas}, which shows results from models with $\h2$ fraction prescription 1. In the top left panel, the black solid curve shows the mean gas-phase metallicity gradient (in units of dex/$r_{90}$) as a function of galaxy stellar mass. The black dashed curves indicate the 1$\sigma$ scatter around the mean. As can be seen, the mean gradient increases from -0.35 dex$/r_{90}$ for galaxies with stellar masses $M_* \sim 10^{9.5} M_{\odot}$ to values near zero for galaxies with $M_* > 10^{10.5} M_{\odot}$. In the top middle panel, the gas-phase metallicity gradient is plotted as function of bulge-to-total ($B/T$) mass fraction of the galaxy. A slightly stronger, and more linear correlation is seen with $B/T$ than with stellar mass. In the top right panel, we plot the average bulge mass fraction as a function of stellar mass, which shows that more massive galaxies tend to have higher bulge mass fractions (and hence weaker metallicity gradients).

In the bottom right panel, the gas-phase metallicity gradient is plotted as a function of $B/T$ for model galaxies in a narrow stellar mass interval ($10.0<\log_{10}[M_*/\ms]<10.2$) and in the bottom left panel metallicity gradient is plotted against $M_*$ in a narrow interval of $B/T$ ($0.25< B/T < 0.3$). These two panels demonstrate that the gas-phase metallicity gradient correlates primarily with bulge mass fraction.

The purple dots superposed on the top three panels of Fig. \ref{fig:mstarzgas} show data from Moran12. We convert the concentration index $r_{90}/r_{50}$ measured for galaxies in this sample to a rough estimate of $B/T$ using the fitting equation in Gadotti (2009) $r_{90}/r_{50}=1.93+2.02~B/T$. Comparing the purple dots with black curves, the qualitative trends are in line with the model predictions, but the scatter is very large -- once again, this is because the data points represent measurements along a 1-dimensional slit, rather than averages in radial bins as in the models. The sample is too small to investigate trends in metallicity gradient as a function of bulge fraction in a narrow stellar mass bin or as a function of stellar mass in a narrow bin of bulge-to-total ratio. It is thus not possible to assess whether the observed gas-phase metallicity gradient is more intrinsically correlated with stellar mass of $B/T$. In addition, we note that there are some galaxies in the observational data set with positively sloped radial metallicity gradients. This is not seen in the models. Integral field spectroscopic observations will be required in order to ascertain whether the outliers with positive gradients are still present in similar numbers once metallicity is averaged radially.

\subsection{Relations between $\boldsymbol Z_{\rm gas}$, sSFR and $\boldsymbol \mu_*$ in the inner and the outer regions of galaxies}

In this section, we will examine correlations between gas-phase metallicity, specific star formation rate and stellar mass surface density. In particular, we will ask whether these relations are the same or are different in the inner and the outer regions of galaxies. In a simple ``closed-box'' model, where gas is transformed into stars at a rate regulated by its density, chemical enrichment proceeds in the same way in the inner and outer regions of the galaxy. The main difference is that less gas is consumed into stars in the outer, lower density regions, resulting in lower gas-phase metallicities. In our semi-analytic models, where the inner regions of disks assemble before the outer regions, and where gas inflows and outflows regulate gas content and metallicity, star formation/metallicity correlations should be very different in the inner and outer regions of galaxies.

Moran12 examined relations between gas-phase metallicity versus specific star formation rate and stellar surface mass density for spectral bins located in the inner ($r<0.7 r_{90}$) and the outer ($r > 0.7 r_{90}$) regions of galaxies. It is difficult to draw clear conclusions
from their analysis, because the individual spectral bin measurements in the outer regions exhibit so much scatter. In this section, we work with SFR-weighted gas-phase metallicities averaged over the inner and the outer disk, which yield considerably less scattered results. We note that the individual gas-phase metallicity measurements are well-constrained, so the scatter is a real physical effect, likely to do with longer timescales for mixing of metals in the outer disk. In our models, heavy elements injected into one radial bin are instantaneously mixed throughout that radial bin, so working with quantities that are averaged over a larger region of the disk should yield a fairer comparison. We restrict our analysis to models that use $\h2$ fraction prescription 1 in this section (very similar results are obtained for the other prescription). The inner disk region is defined as $r<0.7r_{90}$, and the outer disk region as $r>0.7r_{90}$.

\begin{figure*}
\centering
  \includegraphics[angle=-90,scale=0.6]{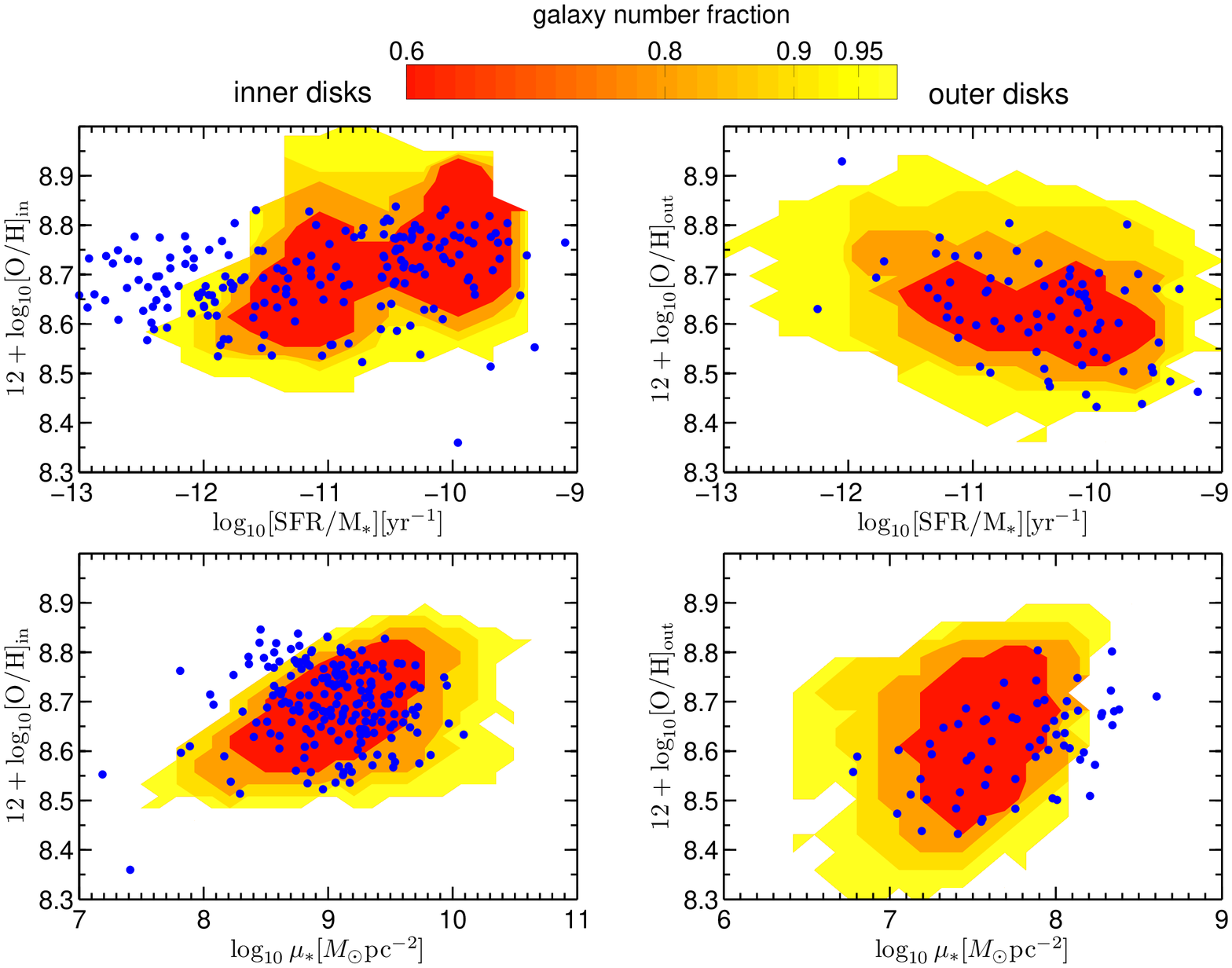}\\
  \caption{The relation of SFR-weighted mean metallicity vs. mean stellar mass-weighted stellar surface density $\mu_*$ and mean specific star formation rate SFR/$M_*$ for inner and outer disks. The left two panels are the results for inner disks and right two panels are for outer disks. The blue dots are from Moran12 data and the contours indicate the fraction of model galaxies located in a given region of parameter space, as given by the colour key at the top of the plot.}
  \label{fig:zgas-ssfr-mustar}
\end{figure*}

In the top panels of Fig. \ref{fig:zgas-ssfr-mustar}, we plot the relation between SFR-weighted gas-phase metallicity and mean specific star formation rate (sSFR=SFR/$M_*$) in the inner disk (left) and in the outer disk (right). The blue dots show results for the Moran12 sample and the red and yellow contours show results for the models. We have taken care to compute the average inner and outer metallicities and sSFR in the same way in the models as in the data.

It is now clear that in the Moran12 sample, the metallicity-sSFR relations are quite different in the inner and the outer regions of the galaxy. At fixed sSFR, the gas-phase metallicity is systematically higher in the inner disk compared to the outer disk. The systematic offset increases towards higher values of SFR/$M_*$. In the inner region of the galaxy, gas-phase metallicity and SFR/$M_*$ are weakly correlated. Regions with the highest specific star formation rates have slightly higher gas-phase metallicities. However, the opposite is true in the outer regions of the galaxy: gas-phase metallicity {\em decreases} at higher specific star formation rates. The data and the models agree quite well. \footnote {We note that the sSFR can no longer be estimated at all accurately below a value of $\sim -12$, so the extension of the data points to sSFR values of $\sim -13$ should not be regarded as a significant discrepancy.}

The reason for the ``inverted'' metallicity-sSFR relation in the outer disk can be clarified by examining the bottom two panels of Fig. \ref{fig:zgas-ssfr-mustar}, where we plot gas-phase metallicity as a function of mean stellar mass-weighted stellar surface mass density
in the inner (left) and outer (right) regions of galaxies for both models and data. Because stellar surface density is a very steeply declining function of radius in disks (top left panels in Fig. \ref{fig:radialprofiles}), the range of stellar surface densities in inner and outer disks is almost completely disjoint. In the data, gas-phase metallicity correlates with stellar surface mass density only in the outer regions of galaxies, where stellar surface densities are low. In the models, there is a correlation between metallicity and $\mu_*$ in both the inner and outer regions, but it is nevertheless very much stronger in the outer disk. Our ``inside-out'' disk formation models predict that most of the stellar mass in the inner disk formed at high redshifts by cooling and collapse of gas within a denser progenitor halo. The outer disks are still in the process of formation at present as the halo continues to accrete dark matter and higher angular momentum gas is able to cool. The outer disks with the lowest stellar surface mass densities are found in galaxies where gas has been accreted, but has not yet reached high enough densities to form molecules and stars. The fact that the outer gas-phase metallicities in models do not reach metallicities much below 0.4 solar arises because gas that cools from the surrounding halo has been significantly pre-enriched with heavy elements (see discussion in Sec. \ref{chap:last}).

\begin{figure*}
\centering
  \includegraphics[angle=-90,scale=0.6]{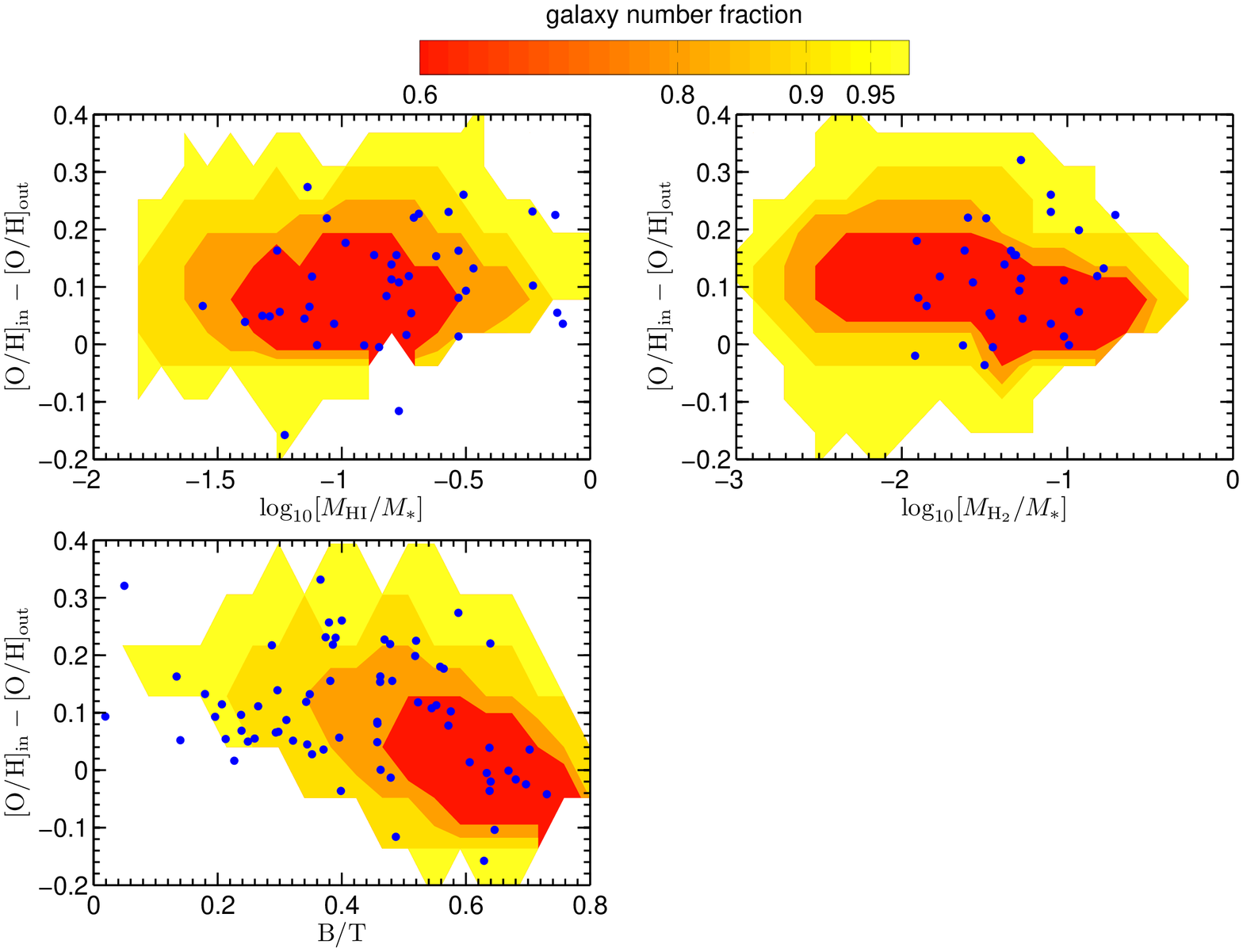}\\
  \caption{The relation of inner and outer disk mean gas-phase metallicity difference vs. total $\Hs$, $\HIs$ and $B/T$ ratios. The meaning of blue points and colour contour are same to those in Fig. \ref{fig:zgas-ssfr-mustar}.}
  \label{fig:drop}
\end{figure*}

Finally, we would like to comment on the result in Moran12 that around $10\%$ of disk galaxies exhibit outer ``metallicity drops'', i.e. they exhibit a sharp turndown in metallicity beyond $\sim r_{90}$. The magnitude of this drop was found to be strongly correlated with total $\HIs$ ratio, but not with total $\Hs$ ratio. We have already commented on the large scatter in metallicity from one spectral bin to another in the outer regions of galaxies in the Moran12 sample. Indeed, close examination of individual profiles in Figure 10 of Moran12 reveals that there is significant diversity in profile shape, even among the ``metal-drop'' objects.

In the top two panels Fig. \ref{fig:drop}, we examine how the inner/outer disk metallicity difference $\rm{[O/H]_{in}-[O/H]_{out}}$ correlates with total $\HIs$ and $\Hs$ ratios. Interestingly, for both the Moran12 data and for our model galaxies the metallicity {\em difference} does not correlate with either $\HIs$ or $\Hs$. This suggests that outer metallicity drops reflect the presence of {\em isolated} spectral bins with low metallicity. Once again integral field spectroscopy data would be extremely useful to test this conjecture in more detail. Inspired by the relation between gas-phase metallicity radial gradient vs. $B/T$ fraction in Sec. \ref{chap:dz-ms}, we plot metallicity difference as a function of $B/T$ in the bottom left panel of Fig. \ref{fig:drop}. This anti-correlation is by far the strongest one in both models and data.

\section{Summary and Discussion} \label{chap:last}

In this paper, we describe how we have transplanted the Fu10 prescriptions for modelling the gas and stellar profiles of disk galaxies
and for tracking the conversion of atomic to molecular gas to the semi-analytic model framework of Guo11. Our models are run on dark matter halo merging trees constructed from both the MS and MS-II. Each galaxy disk is divided into a series of radial concentric rings. This allows us to track the radial distribution of gas, stars and metals in each galaxy.

The main changes with respect to the Fu10 recipes are:

\noindent (i) We adopt a simple star formation law in which the star formation rate surface density is proportional to the molecular gas surface density $\Sigma_{\rm SFR}\propto\Sigma_{\h2}$, rather than the two regime star formation model in Fu10.

\noindent (ii) In Krumholz et al. $\h2$ prescription, we adopt a metallicity-dependent gas clumping factor so that the gas is assumed to be more clumpy in low metallicity galaxy regions. This accelerates molecule formation, star formation and metal production in low mass galaxies and in low metallicity regions and allows us to obtain convergent results in simulations with different resolutions.

\noindent (iii) SNe feedback processes are no longer assumed to be less efficient in the regions with higher gas surface density. Instead, we include radial gas inflow so that cold gas from the outer disk moves inwards to compensate the gas consumption in the inner disk.

\noindent (iv) The prescription for the mixing of heavy elements produced by star formation has been modified. Instead of mixing all the metals directly with the cold gas in the disk, we mix $80\%$ of the metals with the hot gas in the halo, and $20\%$ of the metals with the cold gas in the disk. This partition corresponds roughly to the fraction of total metals produced by SNe as compared to AGB stars.

Based on the model results, we study the radial profiles of gas-phase metallicity and star formation rate surface density and we also examine the correlation between gas-phase metallicity gradient and some global galaxy properties. The main conclusions are:

\noindent (i) The radial gas inflow prevents too fast gas consumption in the inner region of the galaxy disk, because gas flowing in from the outer disk compensates for the consumption. The surface density profiles of molecular gas in $L_*$ galaxies can constrain the inflow velocities.

\noindent (ii) The radial gas inflow has only weak influence on the gas and stellar metallicity profiles, especially {\em in the outer regions of galaxies}. Because of inside-out disk growth, the outer disk metallicity is mainly affected by the gas accreted recently, which has been enriched by star formation and SN feedback in early epochs. On the other hand, the small inflow velocity in inner disk leads to weak influence of inner disk metallicity.

\noindent (iii) The gas-phase metallicity gradient is strongly affected by the fraction of metals directly injected into the halo gas from dying stars, rather than the interstellar cold gas of the galaxy. Metals ejected from a galaxy in early epochs are later re-accreted and this leads to {\em flatter} present-day gas-phase metallicity gradients. We demonstrate that a prescription in which $80\%$ of all the metals are injected into the halo gas provides the best fit to the relatively shallow observed metallicity gradients of galaxies with stellar masses greater than $10^{10} M_{\odot}$ (Kewley et al. 2010; Moran12). We also show that such a prescription results in a good fit to the relation between gas-phase metallicity and specific star formation rate in the outer parts of galactic disks, which are still being built by gas accretion at the present day.

\noindent (iv) Bulge formation through galaxy mergers is the other main process that determines the strength of the gas-phase metallicity gradient. This is because most of the gas in the galaxy is consumed when the bulge is formed in merger induced starburst, and the metallicity gradient is re-established once new gas is able to accrete. In the models, galaxies with the strongest gas-phase metallicity gradients are those that have accreted gas in an undisturbed way over the age of the Universe and that have low bulge mass fractions. We have re-examined metallicity gradient trends in the Moran12 sample and we do indeed find that the gas-phase metallicity gradient correlates more strongly with bulge-to-total ratio than with any other property.

It is worth comparing our results with other recent work on modelling metallicity gradients. Pilkington et al. (2012) use 25 galaxies from four previous hydrodynamical simulation samples (Stinson et al. 2010; Rahimi et al. 2011; Kobayashi \& Nakasato 2011; Few et al. 2012) together with two simple chemical evolution models (Chiappini et al. 2001; Moll{\'a} \& D{\'{\i}}az 2005) to study the metallicity gradients in disks in cosmological context. They study the radial abundance gradient of stars at different age and find that the stellar radial metallicity gradient tends to be flatter at lower redshifts. They conclude that the stellar abundance radial gradient originates from an inside-out disk formation, and the merger histories have less effect on the metallicity gradient than the recipes that treat the sub-grid physics. Gas-phase metallicity gradients are not directly considered in this paper.

Spitoni \& Matteucci (2011) use chemical evolution models that include gas radial inflow processes to study the radial metallicity gradient for Milky Way disk. They test constant inflow and inflow velocity proportional to the galactocentric radius with constant and inside-out infall prescriptions in the models, and they conclude that the most important factor in reproducing the abundance gradient is the radial inflow with a variable speed. They claim that radial inflows are required to explain the observed metallicity gradients in the Milky Way.

The main difference between our model and these models for the origin of radial abundance gradient is the metallicity of the cooling gas.
Most of other models assume the cooling gas always has the initial metallicity, while our model assumes that the infalling gas is pre-enriched by star formation and SN ejecta in early epochs, which leads to different conclusions regarding outer disk metallicities and radial gradients.

\begin{figure}
\centering
 \includegraphics[angle=-90,scale=0.55]{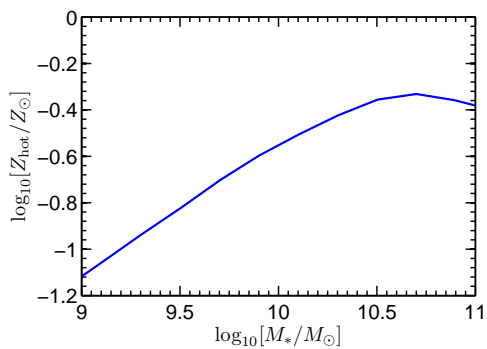}
 \caption
{The mean hot gas metallicity relative to the solar value as a function of the stellar mass of the central galaxy.}\label{fig:zhot}
\end{figure}

In Fig. \ref{fig:zhot}, we plot the metallicity of the hot gas surrounding present-day central galaxies as a function of their stellar mass. This metallicity will be the same as that of the recently accreted gas. As can be seen, the hot gas metallicity is predicted to increase from around 0.1 solar for central galaxies with stellar masses of $\sim 10^9 M_{\odot}$ to around 0.4 solar for galaxies with masses comparable to or greater than that of the Milky Way. We note that significant fraction of the hot gas around central galaxies originates from material that has been ejected out of galaxies by supernovae in early epochs. In low mass haloes, SNe eject most of the gas out of the halo into the so-called ``ejected'' component (see Fig. \ref{fig:processes}). Re-incorporation of this ejected gas into the halo occurs after a few dynamical times for a Milky Way type galaxy, but substantially longer in dwarf systems (see Eq. \ref{eq:gammaej}).

The value of accreted gas metallicity in Fig. \ref{fig:zhot} is consistent with the recent observations by Bresolin et al. (2012), who find that the gas-phase metallicities in the outer disks of some nearby spiral galaxies (NGC 1512, NGC 3621, M83, NGC 4625) are about $0.35Z_{\odot}$. This could reflect the accretion of pre-enriched gas from the intergalactic medium. Our results also agree with earlier work by Tosi (1988), who includes the infall of pre-enriched gas in galaxy chemical evolution models and conclude that the metallicity of infalling gas cannot be higher than $0.4Z_{\odot}$ for Milky Way sized galaxies, otherwise the metallicity gradient in the model will be too flat compared to the Milky Way observations. We note that the gas-phase metallicity gradient of the Milky Way is slightly steeper than typical galaxies of the same stellar mass (left panel of Fig. \ref{fig:inflowz}). In our model, the relatively high metallicity of infalling gas for Milky Way sized galaxy leads to relatively flatter metallicity gradient compared to the observations in Milky Way.

The effect of galaxy merger on the metallicity gradient is also interesting. The observations in galaxy close pairs presented by Kewley et al (2010) show that there is strong relationship between metallicity gradient and galaxy interactions and mergers. Some hydrodynamic simulation works (e.g Rupke, Kewley \& Barnes 2010; Torrey et al. 2012) conclude that the gas inflows by galaxy tidal interaction during galaxy mergers disrupt and flatten the gas-phase metallicity gradient, which is similar to what is predicted by our semi-analytic model. Although we do not trace the detail of gas flows during the galaxy mergers, gas-phase metallicity gradient is disrupted by the merger-induced starburst in our model. As a result, galaxies that have experienced more mergers in their history (or higher $B/T$ ratio) tend to have flatter metallicity gradients.

Finally, we should caution that we have used a very simple model Eq. (\ref{eq:vr}) to describe radial gas inflows and our models also neglect radial transport of gas due to bar instabilities in disks. Some observational constraints on inflows do exist (e.g Levine et al. 2006; Zhang \& Buta 2012), but the measurements are often complicated by non-radial motions in individual galaxies, and average inflow measurements for galaxy samples selected by stellar mass, size, presence or absence of bars etc. are still lacking. In principle, the stellar and gas-phase metallicity profiles in the inner regions of galaxies do provide additional constraints on the inflow prescriptions (see Fig. \ref{fig:inflowz}) and the hope is that future data sets will motivate more detailed work on this topic.

In this paper, we note that the evolution of radial gas, SFR and metallicity gradients to higher redshifts is a topic that is not addressed and that may be of interest to explore in future work.

\section*{Acknowledgments}

We are grateful to the comments from the anonymous referee.

\def\apj{ApJ}
\def\apjl{ApJL}
\def\apjs{ApJS}
\def\aj{AJ}
\def\aap{A\&A}
\def\araa{ARA\&A}
\def\aapss{A\&AS}
\def\mnras{MNRAS}
\def\nature{Nature}
\def\apss{Ap\&SS}
\def\pasp{PASP}

{}

\appendix

\section{The physical processes} \label{chap:appendix}

In this appendix, we will give very brief description for the physical processes related to this paper in the L-Galaxies semi-analytic models of galaxy formation, especially the processes in Fig. \ref{fig:processes}. The following introduction is mainly from model by Guo11 and Guo13, which is based on MS and MS-II halos. As our previous work Fu10 is based DLB07 model, we also mention the equations in DLB07, which is a previous version of the model based on MS haloes.

In both MS and MS-II, the time interval between two successive output snapshots is about 200 Myr. To model the physical processes affecting the baryonic matter, each snapshot is divided into 20 timesteps. Each timestep is about 10 Myr, which is approximately the timescale for massive stars to evolve off the main sequence. All the physcial processes in L-Galaxies are updated in each timestep, but the dark matter halo properties from MS and MS-II are only updated when one snapshot begins.

As discussed in White \& Frenk (1991), the hot gas distributes isothermally in the dark matter halo. When the cooling radius $r_{\rm cool}$ (defined as the radius where the local cooling time equals to the halo dynamical timescale) is smaller than the halo virial radius $r_{\rm vir}$, the halo is in cooling flow regime. The halo hot gas cools to the central galaxy quasi-statically with cooling rate
\begin{equation}\label{eq:mcool}
{\dot m_{{\rm{cool}}}} = \frac{{{m_{{\rm{hot}}}}{r_{{\rm{cool}}}}}}{{{r_{{\rm{vir}}}}t_{{\rm{dyn}}}^{{\rm{halo}}}}}
\end{equation}
in which $t_{\rm dyn}^{\rm halo}=0.1H(z)^{-1}$ is the dynamical timescale of the halo. When $r_{\rm cool}>r_{\rm vir}$, the halo is in the rapid infall regime and all the halo hot gas will accrete on to the central galaxy through a ``cold flow'' in one timestep. The hot gas in the halo is assumed to have specific angular momentum identical to the dark matter halo and the spin parameter of the cooling gas is defined as $\lambda = J{|E|^{1/2}}{G^{-1}}M_{\rm{vir}}^{-5/2}$ according to Mo, Mao \& White (1998), in which $J$ and $E$ are the angular momentum and energy of the dark matter halo.

AGN feedback can suppress the cold gas accretion process. The black hole in the central galaxy can accrete some hot halo gas through the so-called ``radio mode'' and the energy is fed back to the hot halo. Thus it can decrease or even stop the cooling flow to galaxy disks and quench the supply of cold gas for star formation in high mass galaxies. The radio mode accretion rate is
\begin{equation}\label{eq:radio}
{\dot m_{{\rm{BH,R}}}} = {\kappa_{{\rm{AGN}}}}\left( {\frac{{{m_{{\rm{BH}}}}}}{{{{10}^8}{M_ \odot }}}} \right)\left( {\frac{{{f_{{\rm{hot}}}}}}{{0.1}}} \right){\left( {\frac{{{v_{{\rm{vir}}}}}}{{200\rm{km~s}^{-1}}}} \right)^3}
\end{equation}
in which $m_{\rm BH}$ is the black hole mass, $f_{\rm hot}$ is the ratio of hot gas to halo mass, and $v_{\rm vir}$ is the virial velocity of the halo. The power of energy supplied by the black hole accretion is
\begin{equation}\label{eq:EBH}
\dot E_{\rm BH}=0.1\dot m_{\rm{BH,R}}c^2
\end{equation}
in which $\dot m_{\rm{BH,R}}$ is from Eq. (\ref{eq:mcool}) and $c$ is the speed of light. The energy is used to suppress the amount of cooling gas, and thus the cooling rate is
\begin{equation}\label{eq:mcool2}
\dot m{'_{{\rm{cool}}}} = {\dot m_{{\rm{cool}}}} - \frac{\dot E_{\rm BH}}{0.5v_{\rm{vir}}^2}
\end{equation}
for $\dot E_{\rm BH}<0.5v_{\rm{vir}}^2 \dot m_{\rm{cool}}$ and 0 for $\dot E_{\rm BH} \ge 0.5v_{{\rm{vir}}}^2{\dot m_{{\rm{cool}}}}$.

Stars form from the cold gas in ISM, and the star formation model in Guo11 \& DLB07 is
\begin{equation}\label{eq:sfr0}
\dot m_* = \alpha \left( {{m_{{\rm{gas}}}} - {m_{{\rm{crit}}}}} \right)/{t_{{\rm{dyn}}}}
\end{equation}
for cold gas mass $m_{\rm gas}$ larger than critical mass $m_{\rm crit}$, which is a simplified version of star formation law by Kennicutt et al. (1998) including disk instabilities (Toomre 1964). The star formation efficiency $\alpha$ is a constant and the disk dynamical timescale is defined $t_{\rm dyn}=3r_{\rm gas}/v_{\rm max}$ as in Guo11 ($r_{\rm gas}$ is the scale length of gas disk and $v_{\rm max}$ is the maximum circular velocity of the subhalo) and $t_{\rm dyn}=3r_{\rm d}/v_{\rm vir}$ as in DLB07 ($r_{\rm d}$ is the scale length of the galaxy disk and $v_{\rm vir}$ is the virial velocity of the subhalo).

As the stars evolve, a fraction $R$ of the newly formed stars in each timestep is returned instantaneously to the interstellar gas. According to the Chabrier 2003 IMF, $R=0.43$ is adopted. The metal elements from the star formation are all injected into the interstellar cold gas and mixed immediately, and the mass of newly produced metal elements is
\begin{equation}\label{eq:snenergy}
\Delta m_{\rm Z}=Y \Delta m_*
\end{equation}
where $Y=0.03$ is the yield of all metal elements and $\Delta m_*$ is the mass of newly formed stars in a given timestep.

The energy from supernova explosion is
\begin{equation}\label{eq:snenergy}
\Delta E_{\rm SN}=0.5\epsilon_{\rm halo}\Delta m_* v_{\rm SN}^2
\end{equation}
where $0.5v_{\rm SN}^2$ is the energy of supernova ejecta per unit mass of newly formed stars, and $v_{\rm SN}=630~\rm{km~s^{-1}}$ is adopted. $\epsilon_{\rm halo}$ in Eq. (\ref{eq:snenergy}) is a halo-dependent efficiency. The supernova energy can reheat part of the disk cold gas into halo hot gas, and the mass of reheated cold is
\begin{equation}\label{eq:reheat}
\Delta m_{\rm reheat}=\epsilon_{\rm disk}\Delta m_*
\end{equation}
where $\epsilon_{\rm disk}$ is a disk-dependent efficiency. If the supernova energy is large enough, the excess energy of supernova $\Delta E_{\rm SN}-\Delta E_{\rm reheat}$ will eject part of the hot gas out of the halo and be placed in the ejected component. The mass of ejected hot is
\begin{equation}\label{eq:meject}
\Delta m_{\rm eject}=2\left(\Delta E_{\rm SN}-\Delta E_{\rm reheat}\right)/v_{\rm max}^2
\end{equation}
In the previous version of L-Galaxies code DLB07, $\epsilon_{\rm halo}$ and $\epsilon_{\rm disk}$ in supernova feedback are both constants and $\epsilon_{\rm disk}=3.5$, $\epsilon_{\rm halo}=0.35$ are adopted for all galaxies. In Guo11, both $\epsilon_{\rm halo}$ and $\epsilon_{\rm disk}$ are related to the galaxy halo
\begin{equation}\label{eq:epsilonguo}
\begin{array}{l}
\epsilon_{\rm disk}=\epsilon\left[ {0.5 + {{\left( {\frac{{{v_{{\rm{max}}}}}}{{{v_{{\rm{reheat}}}}}}} \right)}^{ - {\beta _1}}}} \right]\\
\epsilon_{\rm halo}=\eta\left[ {0.5 + {{\left( {\frac{{{v_{{\rm{max}}}}}}{{{v_{{\rm{eject}}}}}}} \right)}^{ - {\beta _2}}}} \right]
\end{array}
\end{equation}
in which $\epsilon=4$, $\eta=0.18$, $v_{\rm reheat}=80 \rm{km~s^{-1}}$, $v_{\rm eject}=90 \rm{km~s^{-1}}$ and $\beta_1=\beta_2=3.2$ are adopted in Guo13. This kind of supernova reheating and ejection efficiency that scales primarily with the potential well depth of the host halo results in much higher feedback efficiencies in low mass galaxies, which suppresses the star formation in small galaxies and decreases the number of small galaxies so that it fits the observed stellar mass functions at the low mass end.

With the growth of dark matter halos through dark matter particle accretion and halo mergers, the ejected gas will be reincorporated into the halo of the central galaxy and become the hot gas again. The mass of reincorporated gas at each timestep ($\Delta t$) is
\begin{equation}\label{eq:reincorporation}
\Delta {m_{\rm rein}} = \gamma_{\rm{rein}}{m_{{\rm{ejected}}}}\frac{{\Delta t}}{{t_{{\rm{dyn}}}^{{\rm{halo}}}}}
\end{equation}
in which $m_{\rm eject}$ is the mass of the ejecta reservoir and $t_{\rm dyn}^{\rm halo}$ is the halo dynamical timescale. The reincorporation efficiency $\gamma_{\rm rein}$ in DLB07 is a constant $\gamma_{\rm rein}=0.5$, and $\gamma_{\rm rein}$ in Guo11 is related to the halo virial velocity
\begin{equation}\label{eq:gammaej}
\gamma_{\rm rein}=0.3\left( {\frac{{{v_{{\rm{vir}}}}}}{{220\rm{km~s^{-1}}}}} \right)
\end{equation}
In Eq. (\ref{eq:gammaej}), smaller haloes tend to reincorporate the ejecta reservoir more slowly, which also helps to decrease the number of low mass galaxies in the Guo11 models.

\end{document}